\documentstyle[preprint,aps,graphicx,floats]{revtex}

\begin{document}

\renewcommand{\floatpagefraction}{0.8}

\preprint{$
\begin{array}{l}
\mbox{UB-HET-08-01}\\[-3mm]
\mbox{March~2008} 
\end{array}
$}

\title{Searching for $t\bar t$ Resonances at the Large Hadron Collider}

\author{U.~Baur\footnote{baur@ubhex.physics.buffalo.edu}}
\address{Department of Physics,
State University of New York, \\
Buffalo, NY 14260, USA}
\author{L.H.~Orr\footnote{orr@pas.rochester.edu}}
\address{Dept. of Physics and Astronomy, University of Rochester,\\
Rochester, NY 14627, USA}

\maketitle 

\begin{abstract}
\baselineskip13.pt  
Many new physics models predict resonances with masses in the TeV
range which decay into a pair of top quarks. With its large cross
section, $t\bar t$ production at the Large Hadron Collider (LHC) offers
an excellent opportunity to search for such particles. We present a
detailed study of the discovery potential of the CERN Large Hadron
Collider for Kaluza-Klein (KK) excitations of the gluon in bulk
Randall-Sundrum (RS) models in the $t\bar t\to\ell^\pm\nu b\bar bq\bar q'$
($\ell=e,\,\mu$) final state. We utilize final states with one or two
tagged $b$-quarks, and two, three or four jets (including $b$-jets). Our
calculations take into account the finite resolution of detectors, the
energy loss due to $b$-quark decays, the expected reduced $b$-tagging
efficiency at
large $t\bar t$ invariant masses, and include the background
originating from $Wb\bar b+$~jets, $(Wb+W\bar b)+$~jets, $W+$~jets, and
single top + jets production. We derive semi-realistic $5\sigma$
discovery limits for nine different KK gluon scenarios, and compare
them with those for KK gravitons, and a $Z_H$ boson in the Littlest Higgs
model. We also analyze the capabilities of the LHC experiments to
differentiate between individual KK gluon models and measure the
couplings of KK gluons to quarks. We find that, for the parameters and
models chosen, KK gluons with masses up to about 4~TeV
can be discovered at the LHC. The ability of the LHC to discriminate
between different bulk RS models, and to measure the couplings of the KK
gluons is found to be highly model dependent.
\end{abstract}

\newpage


\tightenlines

\section{Introduction}
\label{sec:one}
The first physics run of the  Large Hadron Collider (LHC) is scheduled
for 2008. Investigating jet, weak boson and top quark
production are the prime goals of the 2008 run. Top pair production at
the LHC, with a cross section which is about two orders of magnitude
larger than at the Fermilab Tevatron, will make it possible to
precisely determine the top quark properties~\cite{Beneke:2000hk}. It also 
offers an excellent opportunity to search for new physics in the early
operational phase of the LHC. Once the LHC reaches design luminosity,
$t\bar t$ production will provide access to new phenomena in the
multi-TeV region. Many extensions of the Standard Model (SM) predict
particles which decay 
into $t\bar t$ pairs, and thus show up as resonances in 
the $t\bar t$ invariant mass, $m(t\bar t)$, distribution. The masses of
these particles are typically in the TeV range. For example,
topcolor~\cite{topcolor,Hill:2002ap} and  Little
Higgs~\cite{Arkani-Hamed:2001nc,azuelos,LH,Schmaltz:2005ky,atlaslh}
models predict weakly coupled new vector bosons, models with extra
dimensions~\cite{Arkani-Hamed:1998rs,Randall:1999ee,anton} can have 
Kaluza-Klein (KK) excitations of the
graviton~\cite{Fitzpatrick:2007qr,Arai:2007ts} the
weak~\cite{anton,McMullen:2001zj,Agashe:2007ki} and the strong gauge
bosons~\cite{Agashe:2006hk,Lillie:2007yh,lillie2,abdel,Ghavri:2006kc,nandi,atlasglu,Burdman:2006gy}
which couple to top quarks, while massive axial vector bosons appear in
torsion gravity models~\cite{Belyaev:2007fn}. Resonances in the $t\bar
t$ channel also occur in technicolor~\cite{Eichten:1994nc,pallin},
chiral color~\cite{chicol} and models with a strong $SU(3)\times SU(3)$
gauge symmetry~\cite{Simmons:1996fz,Choudhury:2007ux}. In some
models~\cite{Fitzpatrick:2007qr,Agashe:2006hk,Lillie:2007yh,lillie2,abdel},
the couplings 
of the new particles to light quarks and gluons is suppressed, and the
$t\bar t$ final state becomes their main discovery channel. For a model
independent approach to search for new physics in $t\bar t$ production,
see Ref.~\cite{Frederix:2007gi}.

Top quarks decay either hadronically, $t\to Wb\to bq\bar q'$
($q,\,q'=u,\,d,\,s,\,c$), or 
semileptonically, $t\to Wb\to b\ell\nu$ ($\ell=e,\,\mu$; decays with
$\tau$ leptons in the final state are ignored here). Pair
production of top quarks thus results in so-called ``di-lepton+jets''
events, $t\bar t\to\ell^\pm\nu_\ell{\ell'}^\mp\nu_{\ell'}b\bar b$,
``lepton+jets'' events, $t\bar t\to\ell^\pm\nu b\bar bq\bar q'$, or
the ``all-hadronic'', $t\bar t\to b\bar b+4$~quarks, final
state. Although the di-lepton+jets channel has the smallest background,
it suffers from a small branching ratio (about 4.7\%). Furthermore, the
two neutrinos in the final state make it impossible to reconstruct the
$t\bar t$ invariant mass or the transverse momentum ($p_T$) of the
individual top quark. The all-hadronic final state has the largest
branching ratio ($\approx 46\%$) but also suffers from a very large
background. The 
lepton+jets channel, finally, has a substantial branching fraction
(about 30\%), while the background is moderate. Since the $t\bar t$
invariant mass can be reconstructed, albeit with a two-fold ambiguity,
it is the premier search channel for new physics in $t\bar t$
production. To identify $t\bar t$ lepton+jets events, the LHC
experiments require an isolated charged lepton, missing transverse
momentum, and at least four isolated hadronic jets. For events with more
than four
jets, the four leading (highest transverse momentum) jets  are selected. 
Of these four jets two have to be tagged as a 
$b$-quarks~\cite{atlastdr,cmstdr}. 

Searching for $t\bar t$ resonances with masses in the TeV region is
challenging for several reasons. For top quark transverse momenta larger 
than a few hundred GeV and $t\bar t$ invariant masses above 1~TeV, the
top quark decay products are highly 
boosted and thus almost collinear. This frequently results in
non-isolated leptons and/or merged or overlapping jets for lepton+jets
and all-hadronic $t\bar t$ events, {\it ie.} the
number of jets may be smaller than the number of final state
quarks. Furthermore, the $b$-tagging efficiency in the TeV region may be
significantly smaller than at low
energies~\cite{atlaslh,Lillie:2007yh,atlasglu}. 

Extending the selection criteria to include topologies with fewer jets
and events with only one tagged $b$-quark is 
an obvious strategy for improving the selection efficiency for very
energetic top quarks. On the other hand, this may significantly increase the
background. In Ref.~\cite{Baur:2007ck} we presented a detailed analysis
of the $t\bar t$ lepton+jets finals states with 2, 3, or 4~jets and one
or two tagged $b$-quarks. We showed that the $\ell\nu+2$~jets and
$\ell\nu+3$~jets final states with one or two $b$-tags significantly
improve the chances for discovering new heavy particles in the $t\bar t$
channel at the LHC, although the background from $W+$~jets and single
top production will be non-negligible in the TeV region, even after
imposing suitable cuts. 

In this paper we derive semi-realistic discovery limits for $t\bar t$
resonances in the lepton+jets final states using the results of
Ref.~\cite{Baur:2007ck}. We consider Kaluza-Klein
excitations of the gluon in representative bulk Randall-Sundrum (RS)
models, in particular those described in Refs.~\cite{lillie2}
and~\cite{abdel}. Taking into account the finite resolution of the LHC
detectors, the energy loss due to $b$-quark decays, the expected reduced
$b$-tagging efficiency at large $t\bar t$ invariant masses, and the
background 
originating from $Wb\bar b\,+$~jets, $(Wb+W\bar b)+$~jets, $W+$~jets, and
single top + jets production, we derive $5\sigma$ discovery limits and
contrast them with those found for KK gravitons in bulk RS
models~\cite{Fitzpatrick:2007qr} and the $Z_H$ boson of the Littlest Higgs
model~\cite{LH}. We also study how well the KK gluons of various
bulk RS models can be discriminated and how well their couplings can be
determined at the LHC. In Sec.~\ref{sec:two} we present a brief overview
of the couplings of the KK gluons we consider and give an outline of our
calculation. Numerical results are presented in
Sec.~\ref{sec:three}. Sec.~\ref{sec:four} contains our conclusions.

\section{Kaluza-Klein gluons: Signal and Background}
\label{sec:two}
We concentrate on the search for the first excited state of the gluon,
$G$, in variants of the RS model with the SM fields propagating in the
bulk. Such models can incorporate Grand Unification of
couplings~\cite{unif}, motivate the flavor hierarchy of fermion
masses~\cite{Huber:2000ie}, and incorporate a dark matter
candidate~\cite{dm}. Bulk RS models with large brane kinetic
terms~\cite{large} or an expanded custodial
symmetry~\cite{Agashe:2006hk,cust1,Djouadi:2006rk} may be able to 
protect the $Zb\bar b$ vertex from large
corrections~\cite{cust1,Djouadi:2006rk,zbb,zbb2}. Specifically, we consider
the KK gluons of
the models of Ref.~\cite{abdel} ($E_i$, $i=1,\dots ,4$), the basic RS
model with the SM in the bulk~\cite{Lillie:2007yh}, models with large
brane kinetic terms with magnitude $\kappa r_{IR}=5$ and $\kappa
r_{IR}=20$, and a model with a $SO(5)\times U(1)_X$ bulk gauge
symmetry~\cite{zbb2}. 

The KK gluons of all models considered here couple uniformly 
to left-handed and right-handed light quarks $q=u,\,d,\,s,\,c$. The
couplings and branching ratios to light, bottom and top quarks, $g^q$, 
$g_L^b=g_L^t$, $g_R^b$, and $g_R^t$, 
and the total width, $\Gamma_G$, in units of the KK gluon mass, $M_G$,
are listed in Table~\ref{tab:one}. They agree with the results given in
Refs.~\cite{lillie2} and~\cite{abdel}. Note that KK gluons do {\sl not}
couple vector-like to the quarks of the third generation.
The partial width for the decay of a
KK gluon into a quark-antiquark pair, $q\bar q$, in the limit $M_G\gg
m_q$, is given by
\begin{equation}
\label{eq:width}
\Gamma(G\to q\bar q)=\frac{M_G}{48\pi}\left(g_L^2+g_R^2\right),
\end{equation}
where $g_L$ ($g_R$) is the coupling of the left-handed (right-handed)
quark to the KK gluon. At tree level, KK gluons do not couple to regular
gluons. 
\begin{table}
\caption{The couplings, branching ratios, and the total width in
units of the mass, 
$\Gamma_G/M_G$, of KK gluons in various bulk RS models. $g_s$ is the
strong coupling constant. In order to calculate $\Gamma_G$ we have
assumed that $\alpha_s=g_s^2/4\pi=0.1$. $N$ is the number of the
additional KK custodial partner quarks in the $SO(5)$ model which are
light enough that $G$ can decay into them. }
\label{tab:one}
\vspace{2mm}
\begin{tabular}{ccccccccc}
Model & $g^q$ & $g_L^b=g_L^t$ & $g_R^b$ & $g_R^t$ & $\sum_q BR(G\to
q\bar q)$ & $BR(G\to b\bar b)$ & $BR(G\to t\bar t)$ & $\Gamma_G/M_G$ \\
\tableline
Basic RS & $-0.2g_s$ & $g_s$ & $-0.2g_s$ & $4g_s$ & $1.7\%$ & $5.7\%$ &
$92.6\%$ & 0.153 \\
$\kappa r_{IR}=5$ & $-0.4g_s$ & $-0.2g_s$ & $-0.4g_s$ & $0.6g_s$ &
68.1\% & 10.6\% & 21.3\% & 0.016 \\
$\kappa r_{IR}=20$ & $-0.8g_s$ & $-0.6g_s$ & $-0.8g_s$ & $-0.2g_s$ &
78.5\% & 15.3\% & 6.1\% & 0.054 \\
$SO(5)$, $N=0$ & $-0.2g_s$ & $2.76g_s$ & $-0.2g_s$ & $0.07g_s$ & 2.0\% &
49.1\% & 48.9\% & 0.130 \\
$SO(5)$, $N=1$ & $-0.2g_s$ & $2.76g_s$ & $-0.2g_s$ & $0.07g_s$ & 0.7\% &
16.0\% & 15.9\% & 0.400 \\
$E_1$ & $-0.2g_s$ & $1.34g_s$ & $0.55g_s$ & $4.9g_s$ & 1.1\% & 7.4\% &
91.4\% & 0.235 \\
$E_2$ & $-0.2g_s$ & $1.34g_s$ & $3.04g_s$ & $4.9g_s$ & 0.9\% & 29.7\% &
69.4\% & 0.310 \\
$E_3$ & $-0.2g_s$ & $1.34g_s$ & $0.55g_s$ & $3.25g_s$ & 2.2\% & 14.2\% &
83.6\% & 0.123 \\
$E_4$ & $-0.2g_s$ & $1.34g_s$ & $3.04g_s$ & $3.25g_s$ & 1.3\% & 46.6\% &
52.1\% & 0.198
\end{tabular}
\end{table}
With the exception of models with a large brane kinetic term, the couplings
of KK gluons to light quarks is suppressed, whereas those to top quarks
are enhanced. In these models, $t\bar t$ production offers the best
chance to discover KK gluons. In models with a large brane kinetic term,
KK gluons may also be visible in di-jet production~\cite{lillie2}. In
all models, except those with a large brane kinetic term, the KK gluons
are fairly broad resonances. 

All cross sections in this paper are computed using
CTEQ6L1~\cite{Pumplin:2002vw} parton distribution functions
(PDFs). For the CTEQ6L1 PDF's, the 
strong coupling constant is evaluated at leading order with
$\alpha_s(M_Z^2)=0.130$. The factorization and renormalization scales
for the calculation of the $t\bar t$ signal are set equal to
$\sqrt{m_t^2+p_T^2(t)}$, where $m_t=173$~GeV is the top quark mass. The
value of the top quark mass chosen is consistent with the most recent
experimental data~\cite{:2007bx}. The choice of factorization and
renormalization scales of the background processes is discussed in more
detail below. The SM
parameters used in all tree-level calculations are~\cite{Mangano:2002ea}
\begin{eqnarray}
\label{eq:input1}
G_{\mu} = 1.16639\times 10^{-5} \; {\rm GeV}^{-2}, & \quad & \\
M_Z = 91.188 \; {\rm GeV}, & \quad & M_W=80.419  \; {\rm GeV}, \\
\label{eq:input2} 
\sin^2\theta_W=1-\left({M^2_W\over M_Z^2}\right), & \quad &
\alpha_{G_\mu} = {\sqrt{2}\over\pi}\,G_F \sin^2\theta_W M_W^2,
\end{eqnarray}
where $G_F$ is the Fermi constant, $M_W$ and $M_Z$ are the $W$ and
$Z$ boson masses, $\theta_W$ is the weak mixing angle, and
$\alpha_{G_\mu}$ is the electromagnetic coupling constant in the $G_\mu$
scheme. 

We calculate the $t\bar t\to\ell\nu b\bar bq\bar q'$ cross
section at leading-order (LO), including the contributions from KK
gluons and all decay correlations, using the helicity spinor technique
described in Ref.~\cite{Kleiss:1988xr}. Top
quark and $W$ decays are treated in the narrow width approximation. We
require that at least one $b$-quark  be tagged and that there are a total
of two, three or four jets in the event. We sum over electron
and muon final states and impose the
following acceptance cuts on lepton+jets events at the
LHC ($pp$ collisions at $\sqrt{s}=14$~TeV):
\begin{eqnarray}\label{eq:cuts1}
p_T(\ell)>20~{\rm GeV}, & \qquad & |\eta(\ell)|<2.5, \\ \label{eq:cuts2}
p_T(j)>30~{\rm GeV}, & \qquad & |\eta(j)|<2.5, \\ \label{eq:cuts3}
p_T(b)>30~{\rm GeV}, & \qquad & |y(b)|<2.5,\\ \label{eq:cuts4}
p\llap/_T>40~{\rm GeV}. & \qquad & 
\end{eqnarray}
Here, $\eta$ ($y$) is the pseudo-rapidity (rapidity), $\ell=e,\,\mu$, 
and $p\llap/_T$ is 
the missing transverse momentum originating from the neutrino in
$t\to b\ell\nu$ which escapes undetected. In addition, we impose an
isolation cut on the charged lepton and jets by requiring the separation in
pseudo-rapidity -- azimuth space to be larger than 
\begin{equation}
\Delta R=[(\Delta\eta)^2+(\Delta\Phi)^2]^{1/2}>0.4. \label{eq:cuts5}
\end{equation}
Light quark jets from $W\to q\bar q'$ and $b$-quark jets are merged if 
\begin{equation}
\Delta R(i,j)<0.4,
\end{equation}
$i,\,j=q,\,q',\,b$. If a $b$-quark jet and a light quark jet merge, their
momenta are combined into a $b$-jet. 

The cuts listed in Eqs.~(\ref{eq:cuts1}) --~(\ref{eq:cuts4}) are
sufficient for the LHC operating at low luminosity, ${\cal L}\leq
10^{33}~{\rm cm^{-2}~s^{-1}}$. They should be tightened somewhat for
luminosities closer to the design luminosity, ${\cal L}=
10^{34}~{\rm cm^{-2}~s^{-1}}$. However, this will have only a small
effect on the cross section in the TeV region on which we concentrate in
this paper. 

We include minimal
detector effects via Gaussian smearing of parton momenta according to
ATLAS~\cite{atlastdr} expectations, and take into account the $b$-jet energy
loss via a parametrized function (for details see
Ref.~\cite{Baur:2007ck}). Charged leptons are assumed to be
detected with an efficiency of $\epsilon_\ell=0.85$. 

At low energies, the LHC experiments are expected to tag $b$-jets with
an efficiency of $\epsilon_b\approx
0.6$~\cite{atlastdr,cmstdr}. However, for very energetic top quarks, the
$b$-tagging efficiency is expected to degrade~\cite{Lillie:2007yh}. As
we shall see in Sec.~\ref{sec:three}, the range of $m(t\bar
t)=2.5-4.0$~TeV will be of interest for KK gluon searches at the
LHC. Preliminary ATLAS studies find that, in this region, $\epsilon_b$ is
about a factor~3 smaller than at low
energies~\cite{atlaslh,atlasglu}. For realistic cross section
estimates, a parametrization of $\epsilon_b$ as a function 
of the $b$-quark energy or $p_T$ is needed. Currently, these do not
exist. Except for low energies, $\epsilon_b$ is known only for a few
selected values of $m(t\bar t)$~\cite{atlaslh,atlasglu}. In the
following we therefore assume a 
constant $b$-tagging efficiency of $\epsilon_b=0.2$. Note that, for
$\epsilon_b=0.2$, the cross section for final states with one $b$-tag is
almost one order of magnitude larger than that for two tagged $b$-quarks.

New particles which decay into a pair of top quarks lead to resonances
in the $t\bar t$ invariant mass distribution and to a Jacobian peak in
the top quark transverse momentum distribution. In the following we
therefore concentrate on these observables. The coupling of KK gluons to
the top quark is reflected also in the $p_T$ distribution of the charged
lepton, which acts as an analyzer of the top
polarization~\cite{Lillie:2007yh}. We do not study the $p_T(\ell)$
distribution here.

Since the neutrino escapes
undetected, $m(t\bar t)$ cannot be directly reconstructed. However,
assuming that the charged lepton and the missing transverse momentum
come from a $W$ boson with a fixed invariant mass $m(\ell\nu)=M_W$, it
is possible to reconstruct the longitudinal momentum of the neutrino,
$p_L(\nu)$, albeit with a twofold ambiguity. In our calculations of the
$m(t\bar t)$ distribution in the lepton+jets final state, we
reconstruct the $t\bar t$ invariant mass using both solutions for
$p_L(\nu)$ with equal weight. Jets are counted and used in the
reconstruction of $m(t\bar t)$ if they 
satisfy Eqs.~(\ref{eq:cuts2}) and~(\ref{eq:cuts3}) after merging. 
The energy loss of the $b$-quarks slightly
distorts the $p\llap/_T$ distribution. As a result, the quadratic
equation for $p_L(\nu)$ does not always have a solution. Events for
which this is the case are discarded in our analysis. This results in a
$\approx 10\%$ reduction of the $t\bar t$ cross section in the $m(t\bar
t)$ distribution. More advanced
algorithms~\cite{Barger:2006hm} improve the reconstruction of the mass
of the new physics signal; however, they have little effect on the shape
of the SM $m(t\bar t)$ distribution.  For the background
processes, the $m(t\bar t)$ distribution is replaced by the
reconstructed $Wb\bar b+m$~jets and $Wbj+m$~jets invariant mass
distribution. 

In order to reconstruct the $t$ or $\bar t$ transverse momentum one has
to correctly assign the $b$ and $\bar b$ momenta to the parent top or
anti-top quark. Since it is impossible to determine the $b$-charge on an
event-by-event basis, and we do only require one $b$-tag in the event,
we combine $p\llap/_T$, $p_T(\ell)$, and the 
transverse momentum of the jet with the smallest
separation from the charged lepton to form the transverse momentum of
the semileptonically decaying top quark. The $p_T$'s of the remaining
jet(s) form the transverse momentum of
the hadronically decaying top\footnote{Alternatively, one could select
the combination of jets which minimizes
$|m(jets)-m_t|$~\cite{Affolder:2000dt}.}. We find that the reconstructed
 and true top quark transverse momentum distributions are virtually
identical except for transverse momenta below 50~GeV
where deviations at the few percent level are observed. 

The main background processes contributing to the $\ell\nu+n$~jet final
states with $n=2,\,3,\,4$ are $Wb\bar b+m$~jets, $(Wb+W\bar b)j+m$~jets,
and $Wjj+m$~jets production, $(t\bar b+\bar tb)+m$~jets, $(t+\bar
t)j+m$~jets production with $t\to b\ell\nu$, and $Wbt$, $Wt$ and $Wjt$
production with $t\to bjj$. For each process,
$m=0,\,1,\,2$, and $j$ represents a light quark or gluon jet, or a
$c$-jet. $Wt$ production only contributes to the 2~jet and 3~jet final
states. The $(Wb+W\bar b)j+m$~jets ($(t+\bar t)j+m$~jets) background 
originates from 
$Wb\bar bj+m$~jets 
($(t\bar b+\bar
tb)+m$~jets) production where one of the $b$-quarks is not detected. We
calculate these processes in the $b$-quark structure function
approximation. We have verified that, for $m=0$, the differential cross
sections for $pp\to Wbj$ ($(t+\bar
t)j$) and $pp\to Wb\bar bj$ ($(t\bar b+\bar tb)j$) where one $b$-jet is not 
detected are very similar. All background cross
sections are consistently calculated at LO. To calculate $pp\to Wb\bar
b+m$~jets and 
$pp\to Wjj+m$~jets we use {\tt ALPGEN}~\cite{Mangano:2002ea}. All other
background processes are calculated using {\tt
MadEvent}~\cite{Maltoni:2002qb}. 

Background processes, such as $pp\to Wjj+m$~jets, where one or two jets
are misidentified as $b$-jets are calculated using a misidentification
probability of $P_{q,g\to b}=P_{j\to b}=1/30$ ($q=u,\,d,\,s$) for light
jets, and $P_{c\to b}=1/10$ for charm quarks. Preliminary ATLAS
studies~\cite{atlasglu,atlaslh} have found these values to be
appropriate in the $t\bar t$ invariant mass region around 3~TeV which is
the range on which this paper concentrates. Ideally, one would like to
know $P_{j\to 
b}$ and $P_{c\to b}$ as functions of the jet transverse
momentum. Unfortunately, these parametrizations are presently not
available.

$Wjj+m$~jets production in {\tt ALPGEN}
includes $c$-jets in the final state. Since 
$P_{c\to b}$ is considerably larger than $P_{q,g\to b}$, 
this underestimates the background from $W+$~charm
production. However, the cross section of $W+$~charm final states is
only a tiny fraction of the full $Wjj+m$~jets rate, resulting in an
error which is much smaller than the uncertainty on the background from
other sources. One can also estimate the $W+$~charm cross section
from that of $pp\to Wb\bar b+m$~jets and $pp\to (Wb+W\bar
b)+m$~jets. For the phase space cuts imposed, quark mass effects are
irrelevant. 
Using the values of $P_{c\to b}$ given in
Refs.~\cite{atlaslh,atlasglu,atlastdr}, we find that 
the $(Wc+W\bar c)j+m$~jets ($Wc\bar c+m$~jets) cross section is 
a factor $2-10$ ($5-100$) smaller than the $(Wb+W\bar b)j+m$~jets
($Wb\bar b+m$~jets) rate for the $p_T$ and invariant mass range
considered here.

$b\bar b+m$~jets production where one $b$-quark decays semileptonically
also contributes to the background. 
Once a lepton isolation cut has been imposed, this background is known to
be small for standard lepton+jets cuts~\cite{Hubaut:2005er}. For $b\bar
b+m$~jets events to mimic $t\bar t$ production with very energetic top
quarks, the $b$-quarks also have to be very energetic. This will make
the lepton isolation cut even more efficient. We therefore ignore the
$b\bar b+m$~jets background here. 

The
renormalization and factorization scales,
$\mu_r$ and $\mu_f$,
of background processes
involving top quarks are set to $m_t$; for all other background
processes we choose the $W$ mass. Since our calculations are performed
at tree level, the cross section of many background processes exhibits a
considerable scale dependence. However, uncertainties on the current
$b$-tagging efficiencies and the light jet mistag probability at high
energies introduce an uncertainty which may well be larger.
Our choice of $\mu_r$ and $\mu_f$ leads to a rather conservative
estimate of the background cross sections; other (reasonable) choices such
as $\mu_r^2=\mu_f^2=M_W^2+\sum_i p_T(j_i)^2$, where $i$ runs over all
jets, lead to smaller cross sections, especially at high energies.

Without further cuts, the background turns out to be much larger than
the signal for $t\bar t$ invariant masses in the TeV
region~\cite{Baur:2007ck}. The signal to background ratio, however, can
be improved significantly by imposing a cut
\begin{equation}
\label{eq:mt}
|m_T(j_{min}\ell)-m_t|< 20~{\rm GeV}
\end{equation}
on the cluster transverse mass, $m_T$, and a cut 

\begin{equation}
\label{eq:invm}
|m(t\to nj)-m_t|<20~{\rm GeV},
\end{equation}
on invariant mass of the $n=1,\,2$ or~3 remaining jets which are assumed
to originate from the
hadronically decaying top quark. The cluster transverse mass in
Eq.~(\ref{eq:mt}) is defined by
\begin{equation}
\label{eq:mtcl1}
m^2_{T}(j_{min}\ell)=\left(\sqrt{p_T^2(j_{min}\ell)+m^2(j_{min}\ell)}
+ p\llap/_T\right)^2-\left(\vec{p}_T(j_{min}\ell)+
\vec{p\llap/}_T\right)^2\, ,
\end{equation}
where $p_T(j_{min}\ell)$ and $m(j_{min}\ell)$ are the transverse
momentum and invariant mass of the $j_{min}\ell$ system, respectively,
and $j_{min}$ is the jet with the smallest separation from the charged
lepton. $m_T$ sharply peaks at the top mass. The invariant mass
resolution for jet systems with a mass near $m_t$ is 
approximately $7-10$~GeV for jets with energies above 200~GeV. The
invariant mass window chosen in 
Eq.~(\ref{eq:invm}) thus will capture most of the $t\bar t$ signal. On the
other hand, it is sufficiently narrow to reject a large portion of the
background. 

In case of only two jets in the final state, we impose
Eq.~(\ref{eq:invm}) on the jet with the larger separation from the
charged lepton. In order to estimate the effect of a jet invariant mass
cut on the $Wjj$ and $(t+\bar t)j$ background, we convolute the
differential cross sections obtained from {\tt ALPGEN} and {\tt MadEvent} 
with ${\cal P}(m(j_{top}),p_T(j_{top}))$ where $j_{top}$ is the jet with
the larger separation from the charged lepton ({\it ie.} the ``t-jet''
candidate)~\cite{Baur:2007ck}. A cut on $m(j_{top})$ is then imposed
(see below). ${\cal 
P}(m(j),p_T(j))$ is the two-dimensional probability density that a jet
with transverse momentum $p_T(j)$ has an 
invariant mass $m(j)$. We calculate ${\cal P}(m(j),p_T(j))$ by
generating $10^5$ $W+$~jets events in {\tt
PYTHIA}~\cite{Sjostrand:2006za} and passing them through {\tt
PGS4}~\cite{conway}, which simulates the response of a generic
high-energy physics collider detector with a tracking system,
electromagnetic and hadronic calorimetry, and muon system. Jets are
reconstructed in the 
cone~\cite{cone} and $k_T$ algorithms~\cite{Ellis:1993tq} as implemented
in {\tt PGS4}, 
using a cone size ($D$ parameter) of $R=0.5$ ($D=0.5$) in the cone
($k_T$) algorithm. Since it is infrared safe, the $k_T$ algorithm is the
theoretically preferred algorithm. For a discussion of the advantages
and disadvantages of the two algorithms at hadron colliders, see
Ref.~\cite{Blazey:2000qt}. The cone size ($D$ parameter) is deliberately
chosen to be slightly larger than in our parton level studies to avoid
drawing conclusions which are too optimistic. The probability density
function, ${\cal P}$, for the $k_T$ algorithm has 
a much longer tail at large jet invariant masses than for the cone
algorithm, resulting in a significantly higher background in the
$m(t\bar t)$ distribution~\cite{Baur:2007ck}. In order to be
conservative, we therefore use the $k_T$ algorithm when estimating the
background in the $\ell\nu+2$~jets final state. 

\begin{figure}[th!] 
\begin{center}
\includegraphics[width=13.2cm]{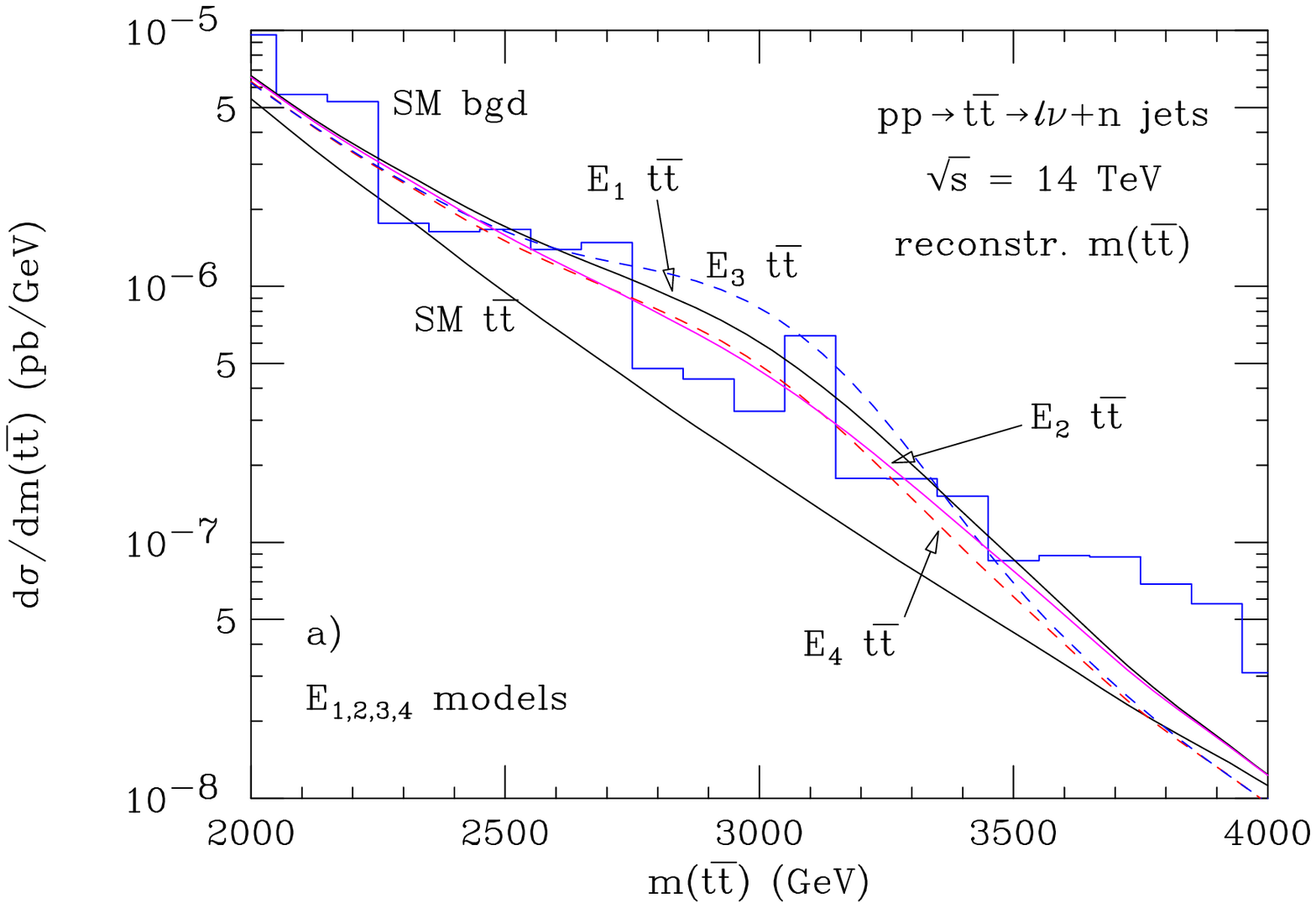} \\[3mm]
\includegraphics[width=13.2cm]{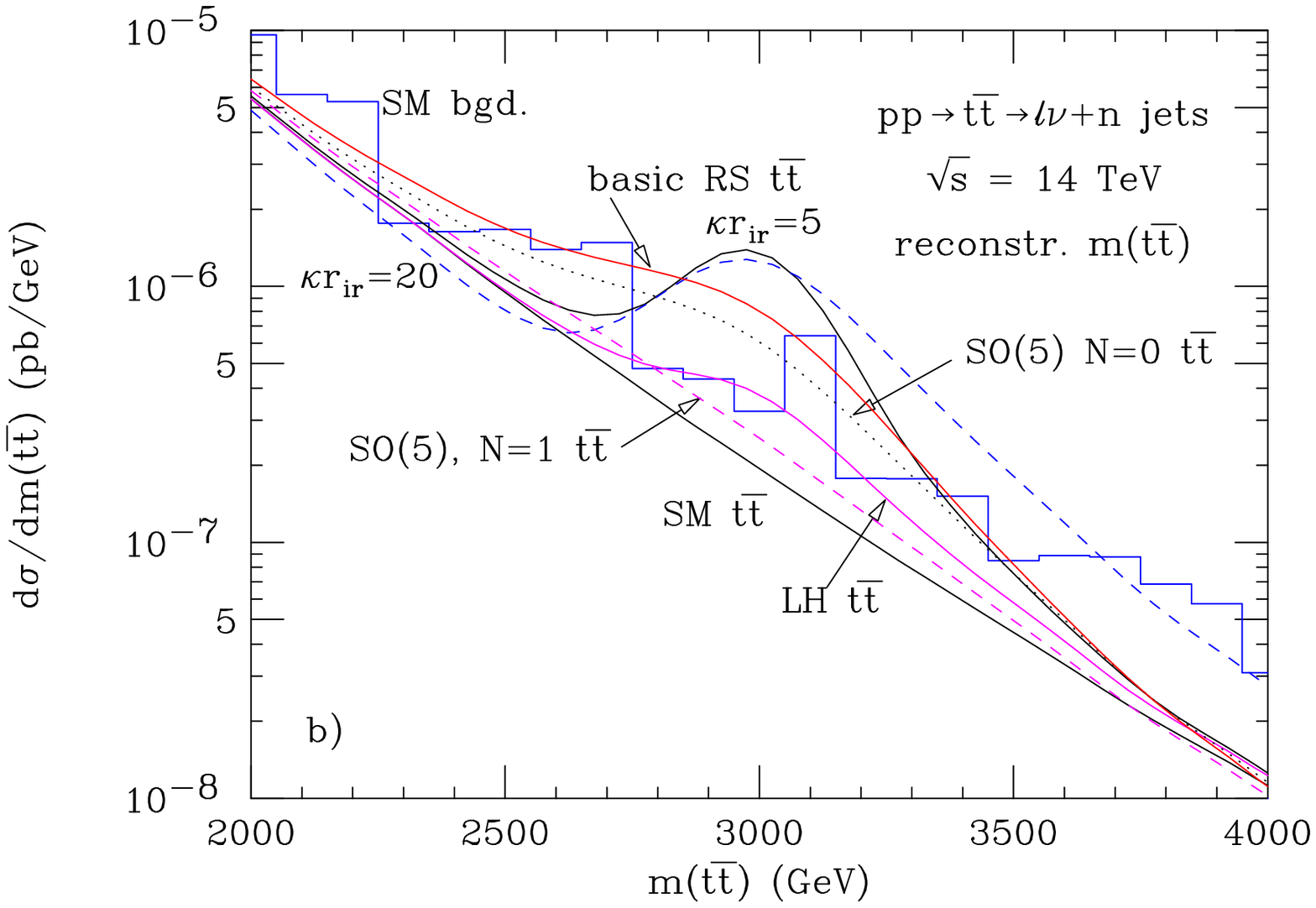}
\vspace*{2mm}
\caption[]{\label{fig:one} 
The LO differential cross section of the combined 
SM $t\bar t\to\ell\nu+n$~jets ($n=2,\,3,\,4$) signal (black line), the
combined background (blue histogram), and a
bulk RS KK gluon, $G$, with $M_G=3$~TeV as a 
function of the reconstructed $t\bar t$ invariant mass. 
One or two of the jets are assumed to be $b$-tagged. Part a) of the
figure shows the results for $E_i$ ($i=1,\dots,4$) KK gluons, part b)
shows the resonance curves for the remaining KK gluon scenarios
summarized in Table~\ref{tab:one}. For comparison, the magenta line in b)
shows the result for a $Z_H$ boson in the Littlest Higgs model with a
mass of 3~TeV and  
$\cot\theta=1$ (see Ref.~\cite{LH}). The cuts imposed are discussed in the
text.} 
\vspace{-7mm}
\end{center}
\end{figure}

\begin{figure}[th!] 
\begin{center}
\includegraphics[width=12.7cm]{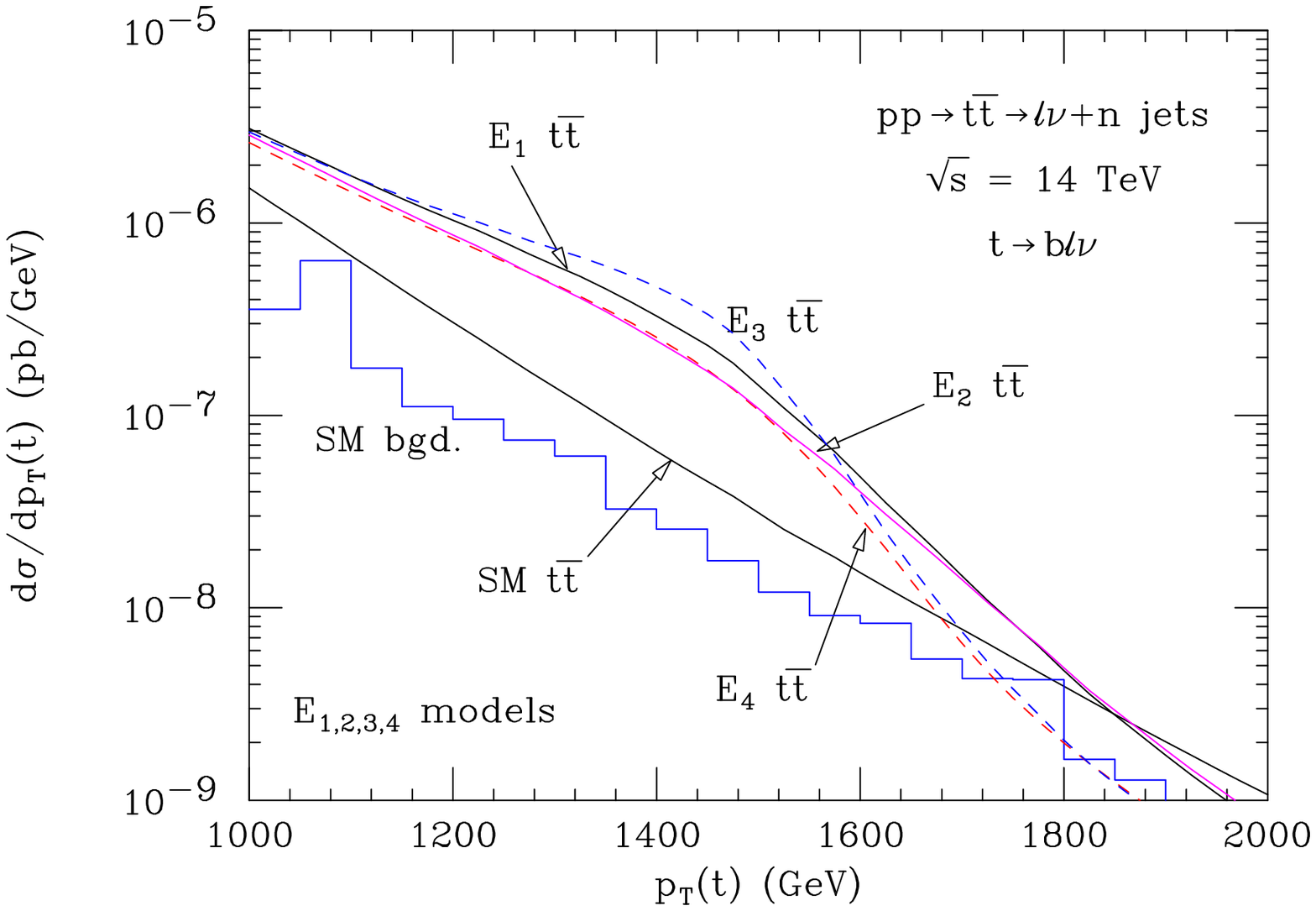} \\[3mm]
\includegraphics[width=12.7cm]{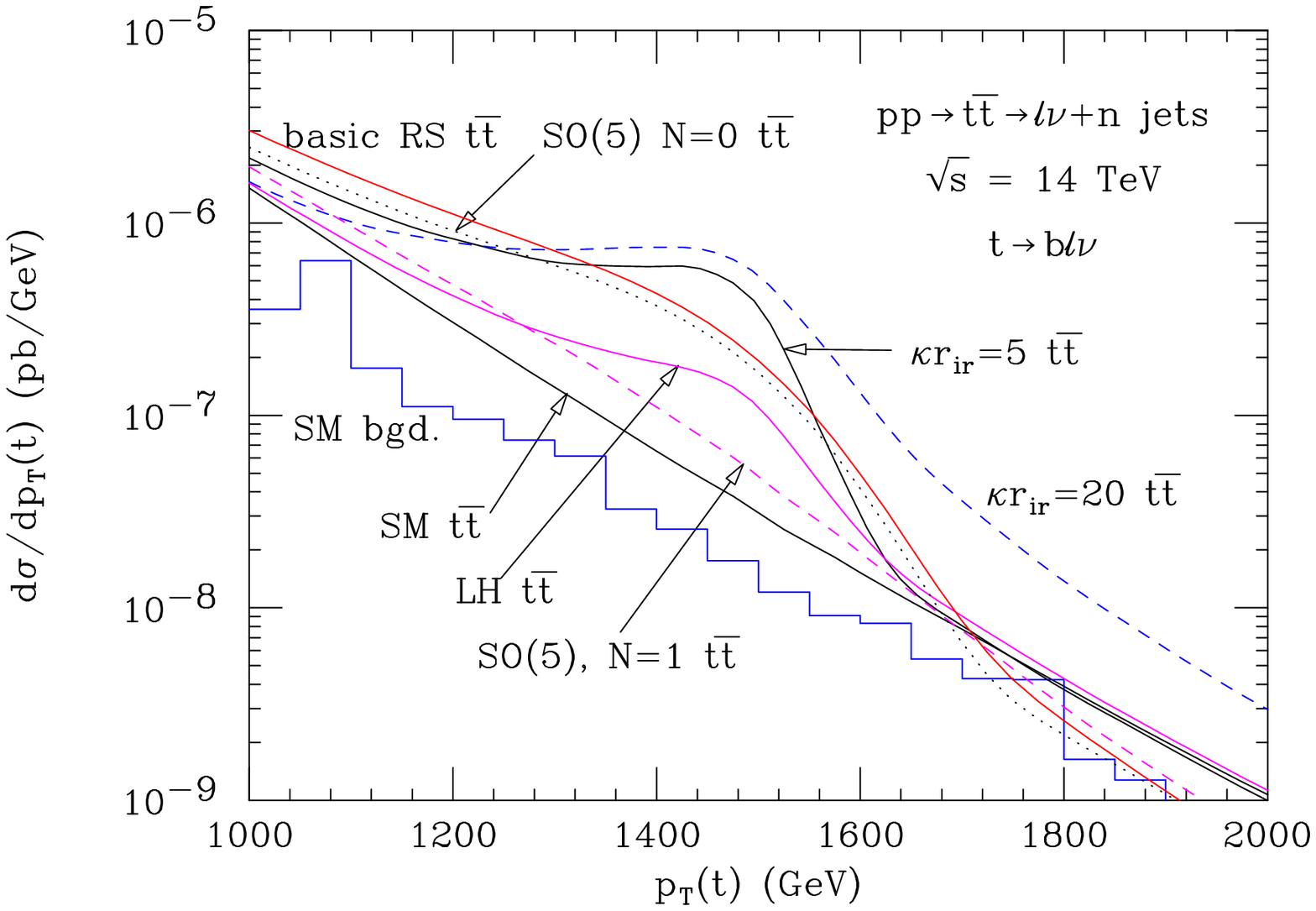}
\vspace*{2mm}
\caption[]{\label{fig:two} 
The LO differential cross section of the combined 
SM $t\bar t\to\ell\nu+n$~jets ($n=2,\,3,\,4$) signal (black line), the
combined background (blue histogram), and a
bulk RS KK gluon, $G$, with $M_G=3$~TeV as a 
function of the reconstructed transverse momentum of the
semileptonically decaying top quark. 
One or two of the jets are assumed to be $b$-tagged. Part a) of the
figure shows the results for $E_i$ ($i=1,\dots,4$) KK gluons, part b)
shows the resonance curves for the remaining KK gluon scenarios
summarized in Table~\ref{tab:one}. For comparison, the magenta line in b)
shows the result for a $Z_H$ boson in the Littlest Higgs model with a
mass of 3~TeV and  
$\cot\theta=1$ (see Ref.~\cite{LH}). The cuts imposed are discussed in the
text.} 
\vspace{-7mm}
\end{center}
\end{figure}

The reconstructed $m(t\bar t)$ distribution after imposing the cuts
listed in Eqs.~(\ref{eq:cuts1}) --~(\ref{eq:invm}) is shown in
Fig.~\ref{fig:one} for the range $|M_G-m(t\bar t)|\leq 1$~TeV. We show
the results for the combined $t\bar t\to\ell\nu +
n$~jets final states with $n=2,\,3,\,4$ and one or two tagged
$b$-quarks, assuming $\epsilon_b=0.2$ and $\epsilon_\ell=0.85$. The
curves are for SM $t\bar t$ production
(solid black line), the combined background (blue histogram), and KK
gluon production with $M_G=3$~TeV for the models listed in
Table~\ref{tab:one}. 
The transverse momentum distribution of the semileptonically decaying
top quark is shown in Fig.~\ref{fig:two}. To avoid
overburdening the figures, we show the $E_i$ ($i=1,\dots, 4$) KK gluon
resonances in Fig.~\ref{fig:one}a and Fig.~\ref{fig:two}a, and all others in
Fig.~\ref{fig:one}b and Fig.~\ref{fig:two}b, respectively.
For comparison, the magenta line in Figs.~\ref{fig:one}b and~\ref{fig:two}b
shows the result for a $Z_H$ boson in the Littlest Higgs model with
$M_{Z_H}=3$~TeV and 
$\cot\theta=1$ (see Ref.~\cite{LH}), where $\theta$ is a mixing
angle. The $Z_H$ vector boson couples purely left-handed and universally
to quarks and leptons.

For all models, except that with $\kappa r_{IR}=20$, the sign of the
coupling of the KK gluon to light quarks is opposite to that of the
larger coupling to the top quark. As a result, in those cases,
interference effects are positive (negative) below (above) the
resonance. Since the width of KK gluons in models with a large brane
kinetic term is relatively small, the resonance curves for these particles
are significantly more pronounced than those for other KK
gluons. For such rather narrow resonances detector resolution effects
become important. These effects are included in Figs.~\ref{fig:one}
and~\ref{fig:two} through the smearing of particle momenta according to
the ATLAS resolution. 

As evident from Fig.~\ref{fig:two}, the SM non-$t\bar t$ 
background is significantly smaller in the $p_T(t\to b\ell\nu)$
distribution than in the $t\bar t$ invariant mass spectrum. Furthermore,
the top quark transverse momentum distribution does not suffer from the
ambiguity associated with the reconstruction of the longitudinal
momentum of the neutrino. On the other hand, the transverse momentum
distribution only reflects information encoded in the transverse degrees
of freedom. 

\section{Numerical Results}
\label{sec:three}

We now derive discovery limits for the KK gluon states discussed in
Sec.~\ref{sec:two}. We also investigate, for $M_G=3$~TeV, how well the
KK gluon states can be discriminated, and how well the couplings of
these states can be measured at the LHC, and a luminosity upgraded LHC
(SLHC) with a total integrated luminosity of 3000~fb$^{-1}$. 

As the statistical tool of choice we adopt a log likelihood test. 
Our expression for the log-likelihood function is
\begin{eqnarray}
\nonumber
-2\log L & = &
-2\left[\sum_i\left(-f_SS_i-f_BB_i+n_{0i}\log(f_SS_i 
                    +f_BB_i) - \log(n_{0i}!)\right)\right]
\\ 
& & + {(f_S-1)^2\over(\Delta f_S)^2}
    + {(f_B-1)^2\over(\Delta f_B)^2} \, .
\end{eqnarray}
The sum extends over the number of bins, $S_i$ and $B_i$ are the
number of signal and background events in the $i$th bin, and $n_{0i}$
is the number of reference (eg. SM) events in the $i$th bin.  The
uncertainties on the 
signal and background normalizations are taken into account via two
multiplicative factors, $f_S$ and $f_B$, which are allowed to vary but
are constrained within the relative uncertainties of the signal and
background cross sections, $\Delta f_S$ and $\Delta f_B$,
respectively. The background consists of SM $t\bar t$ production, and SM
non-$t\bar t$ background as discussed in Sec.~\ref{sec:two}.

Since both the $m(t\bar t)$ and the $p_T(t\to b\ell\nu)$
distributions have advantages and disadvantages, we use both in
deriving discovery and sensitivity limits for the couplings of KK
gluons. As we do not take into account common systematic
uncertainties in our analysis, this procedure will lead to somewhat
optimistic limits. If we use only the distribution which yields the
tightest individual bounds, the results presented in this
Section worsen by $10-20\%$. 

Except for the SM $t\bar t$ cross section, and the backgrounds
contributing to the $\ell\nu+2$~jets final states, cross sections are
only known to leading order in QCD and thus depend significantly on the
renormalization and factorization scales used. In the following, we
assume that QCD corrections do not significantly change the shape of the
distributions analyzed in the region which contributes most to the
statistical significance. Furthermore, we assume that the uncertainties
for signal and background from the unknown QCD corrections are
approximately equal, $f_S=f_B=f$. In this case,  $\log L$ can be
minimized analytically and one finds  the minimum of $\log L$ to occur
at
\begin{equation}
f=\frac{1}{2}\left(
1-(\Delta f)^2N+\sqrt{(1-(\Delta f)^2N)^2+4(\Delta f)^2N_0}
\right) \, ,
\end{equation}
where
\begin{equation}
N=\sum_i(S_i+B_i)
\end{equation}
is the total number of events, 
\begin{equation}
N_0=\sum_i n_{0i}
\end{equation}
the total number of reference events, and $\Delta f$ is the 
uncertainty of the reference cross section. In the following we take
$\Delta f=0.3$. The results which 
we present below only minimally depend on the choice of $\Delta f$,
reflecting that the normalization of the background can be obtained from the
low energy part of the $m(t\bar t)$ and $p_T(t\to b\ell\nu)$
distributions. Uncertainties from parton distribution functions, and
from varying the factorization and renormalization scales are ignored in
our calculation. Uncertainties from the poorly known $b$-tagging
efficiency and light quark/gluon jet misidentification probability are
likely to be larger and difficult to quantify without actual LHC data or
more accurate simulations.

\subsection{Discovery limits}

In order to derive discovery limits for the KK gluons introduced in
Sec.~\ref{sec:two} at the LHC we require a 5~standard deviation significance 
\begin{equation}
-2\log L\geq 25
\end{equation}
from the
SM prediction in the combined reconstructed $t\bar t$ invariant mass, and 
$p_T(t\to b\ell\nu)$ distribution. Results for 100~fb$^{-1}$ and
300~fb$^{-1}$ of data are shown in Table~\ref{tab:two}.  
\begin{table}
\caption{Approximate $5\sigma$ discovery limits for the KK gluons
introduced in 
Sec.~\ref{sec:two} for 100~fb$^{-1}$ and 300~fb$^{-1}$ of data at the
LHC. For comparison, we also show discovery limits for a $Z_H$ boson in
the Littlest Higgs model with $\cot\theta=1$, and a bulk RS KK graviton
with $M_4L=1$ and $\nu_{t,R}=1$. See text for more details. }
\label{tab:two}
\vspace{2mm}
\begin{tabular}{ccc|ccc}
Model & limit 100~fb$^{-1}$ & limit 300~fb$^{-1}$ & Model & limit
100~fb$^{-1}$ & limit 300~fb$^{-1}$ \\ 
\tableline
Basic RS & 3.8~TeV & 4.3~TeV & $E_1$ & 3.9~TeV & 4.4~TeV\\
$\kappa r_{IR}=5$ & 3.4~TeV & 3.9~TeV & $E_2$ & 3.6~TeV & 4.2~TeV\\
$\kappa r_{IR}=20$ & 3.5~TeV & 4.1~TeV & $E_3$ & 3.8~TeV & 4.2~TeV\\
$SO(5)$, $N=0$ & 3.4~TeV & 4.0~TeV & $E_4$ & 3.4~TeV & 4.2~TeV \\
$SO(5)$, $N=1$ & 2.4~TeV & 3.0~TeV & $Z_H$ & 2.6~TeV & 2.8~TeV \\
KK graviton & 1.3~TeV & 1.4~TeV & & &  
\end{tabular}
\end{table}
For comparison, we also list the $5\sigma$ discovery limits for a $Z_H$
boson in the Littlest Higgs model with $\cot\theta=1$, and for a bulk RS
KK graviton, $G_r$, which is dominantly produced via gluon
fusion~\cite{Fitzpatrick:2007qr}. 

To calculate the cross section for $G_r$ production via
gluon fusion in the $t\bar t$ channel we use the formulas of
Ref.~\cite{Fitzpatrick:2007qr} with $M_4L=1$ and $\nu_{t,R}=1$. Here
$M_4$ is the Planck scale, $L$ is the inverse of the AdS curvature
scale, and $\nu_{t,R}$ is a parameter related to the bulk mass for
fermion fields. The $gg\to G_r\to t\bar t$ cross section scales like
$(M_4L)^{-4}$ and $(1+2\nu_{t,R})^2$. 

The discovery limits for KK gluons in all models considered are, except
for the
$SO(5)$ model with $N=1$, in the range $3.4-3.9$~TeV ($3.9-4.4$~TeV)
for 100~fb$^{-1}$ (300~fb$^{-1}$). In the $SO(5)$ model with $N=1$, the
first KK excitations of the fermions are assumed to be sufficiently
light so that KK gluons can decay into those. As a result, the KK gluon
in this model is a very broad resonance (see Table~\ref{tab:one}) which
makes it considerably more difficult to detect. Our discovery limits for the
basic RS case are in general agreement with those obtained in
Ref.~\cite{Agashe:2006hk}. Note that KK gluons in models with a large
bulk kinetic term ($\kappa r_{IR}=5$ and $\kappa r_{IR}=20$) couple more
strongly to light quarks than top quarks and thus can be searched also
for in di-jet production~\cite{lillie2}; however, no quantitative
discovery limits for this channel have been derived yet. Precision
electroweak data allow KK gluons with mass as low as
$2-3$~TeV~\cite{Carena:2006bn}. The LHC thus should be able to
significantly constrain bulk RS models.

We do not list discovery limits for an upgraded LHC with 10~times the
integrated luminosity of the LHC (SLHC). Using a $b$-tagging efficiency
of $\epsilon_b=0.2$, which is appropriate for $t\bar t$ invariant
masses of ${\cal O}(3$~TeV), we obtain $5\sigma$ limits of $M_G>5$~TeV
for KK gluons 
with an integrated luminosity of 3000~fb$^{-1}$. However, some caution
is in order because a $b$-tagging 
efficiency of $\epsilon_b=0.2$ may well be too optimistic at such huge
invariant masses. Unfortunately, currently no estimates exist for
$\epsilon_b$ at the SLHC in the vicinity of $m(t\bar t)=5$~TeV.

The $Z_H$ boson couples with weak coupling
strength to fermions. It is therefore not surprising that the discovery
limits for a $Z_H$ boson in the $t\bar t$ channel are substantially
weaker than those for most KK gluons. Since the $Z_H$ boson also couples
to charged leptons, the $\ell^+\ell^-$ final state is an obvious channel
to search for such a particle. It should be possible to find a $Z_H$
with mass up to 5~TeV in di-lepton production at the LHC with
300~fb$^{-1}$~\cite{azuelos}. 

The discovery limits for bulk RS KK gravitons are about a factor~3
weaker than those 
for KK gluons due to the strongly suppressed $G_rgg$ coupling.
However, the limits listed for  bulk RS KK gravitons in
Table~\ref{tab:two} are likely conservative. The $b$-tagging efficiency
in the $t\bar t$ invariant mass range of $1-1.5$~TeV is estimated to be
a factor $1.5-2$ higher than what we have used in our
calculation~\cite{atlaslh,atlasglu}. This will increase the $5\sigma$
discovery limits for bulk RS KK gravitons by approximately
$100-200$~GeV. For $\nu_{t,R}<1$, the $ZZ$~\cite{Agashe:2007zd} and
$WW$~\cite{Antipin:2007pi} channels may offer better chances to discover
bulk RS KK gravitons. 

In Ref.~\cite{Fitzpatrick:2007qr} $G_r$ discovery
limits were derived as a function of the top quark detection efficiency,
without 
taking into account the non-$t\bar t$ background. Our calculation
attempts to provide a more quantitative estimate, taking into account
the non-$t\bar t$ background, and the reduced $b$-tagging efficiency at
large invariant masses.

\subsection{Discriminating KK gluon models}

Once a resonance in the $t\bar t$ channel has been discovered, it becomes
important to determine its properties in order to pin down the
underlying new physics. The spin of the new particle can be determined
by measuring the angular distribution of the top quarks: a scalar
particle leads to an isotropic distribution, a vector boson to a
distribution which is proportional to $(1+\cos^2\theta)$, whereas the
angular distribution for spin~2 particle will have a $(1-\cos^4\theta)$
dependence~\cite{Fitzpatrick:2007qr,Allanach:2000nr}. Here, $\theta$ is
the scattering angle of the top quark. Important clues
can also be obtained from other final states in which the same resonance
has been observed. 

For the following discussion we assume that a spin~1 resonance has been
found in the $t\bar t$ channel, however, has not 
been observed elsewhere. In such a situation, KK gluons in bulk RS
models become natural candidates for the state observed and it becomes
interesting whether a measurement of the resonance curve in the $m(t\bar
t)$ and the $p_T(t\to b\ell\nu)$ distribution will be able to
discriminate between different bulk RS models. 

In order to address this question, we pursue two approaches. In this
Section, we calculate the ``discrimination matrix'' for KK gluons
in the nine bulk RS models we are considering. In
Section~\ref{sec:threec}, we derive 68.3\% confidence level (CL) bounds for
the couplings of KK gluons. For our case study, we assume a mass of
$M_G=3$~TeV for the KK gluon. This guarantees that the LHC will be able
to detect such a particle with a significance of more than $5\sigma$ in
all models studied here, except the $SO(5)$, $N=1$ case.

The discrimination matrix is constructed by performing a log likelihood
test for each pair of bulk RS models, assuming that one is correct and
finding the significance of the other model as a test. The results for
$M_G=3$~TeV and an integrated luminosity of 100~fb$^{-1}$ are presented
in Table~\ref{tab:three}.
\begin{table}
\caption{Discrimination matrix for the KK gluons introduced in 
Sec.~\ref{sec:two} for $M_G=3$~TeV and an integrated luminosity of
100~fb$^{-1}$ at the LHC. The model in each column is assumed to be the
correct, measured model, and is tested against the hypothesis in each
row. Since the discrimination matrix is symmetric
in the limit of large statistics, we only show the entries above the
diagonal. } 
\label{tab:three}
\vspace{2mm}
\begin{tabular}{c|ccccccccc}
Model & basic RS & $\kappa r_{IR}=5$ & $\kappa r_{IR}=20$ & $SO(5)$,
$N=0$ & $SO(5)$, $N=1$ & $E_1$ & $E_2$ & $E_3$ & $E_4$ \\
\tableline
basic RS & $0.0\sigma$ & $6.1\sigma$ & $10.5\sigma$ & $3.1\sigma$ &
$7.5\sigma$ & $1.7\sigma$ & $2.9\sigma$ & $0.0\sigma$ & $2.9\sigma$ \\
$\kappa r_{IR}=5$ & & $0.0\sigma$ & $5.2\sigma$ & $4.3\sigma$ &
$7.3\sigma$ & $7.1\sigma$ & $7.2\sigma$ & $5.4\sigma$ & $6.0\sigma$ \\
$\kappa r_{IR}=20$ & & & $0.0\sigma$ & $7.8\sigma$ & $9.9\sigma$ &
$10.9\sigma$ & $10.9\sigma$ & $9.3\sigma$ & $9.7\sigma$ \\
$SO(5)$, $N=0$ & & & & $0.0\sigma$ & $4.8\sigma$ & $3.7\sigma$ &
$3.3\sigma$ & $2.8\sigma$ & $1.9\sigma$ \\
$SO(5)$, $N=1$ & & & & & $0.0\sigma$ & $6.8\sigma$ & $5.5\sigma$ &
$6.8\sigma$ & $4.3\sigma$ \\
$E_1$ & & & & & & $0.0\sigma$ & $1.5\sigma$ & $2.0\sigma$ & $2.7\sigma$
\\ 
$E_2$ & & & & & & & $0.0\sigma$ & $3.2\sigma$ & $1.6\sigma$ \\
$E_3$ & & & & & & & & $0.0\sigma$ & $3.2\sigma$ \\
$E_4$ & & & & & & & & & $0.0\sigma$ \\
\end{tabular}
\end{table}
For smaller (larger) KK gluon masses higher (lower) significances are
expected. 

Table~\ref{tab:three} shows that the 
$E_i$ models can only be distinguished at the $1.5-3\sigma$
level. However, the remaining models can be discriminated with a
significance of $4-10\sigma$. The $E_i$ models and non-$E_i$ models,
finally, can be separated at the $2-11\sigma$ level, except for the
basic RS and the $E_3$ model which will be very hard to discriminate
through a measurement of the resonance curve for the 
mass and the integrated luminosity chosen. This
can be easily understood. At the resonance peak, $m(t\bar t)=M_G$, the
$q\bar q\to G\to 
t\bar t$ cross section is proportional to $Br(G\to q\bar q)\cdot Br(G\to
t\bar t)$, where $q=u,\,d,\,s,\,c$ denotes a light quark. For a KK gluon
in the basic RS and the $E_3$ model, the product of the two branching
fractions accidentally agrees within 15\%, making it very difficult to
discriminate between the two models. Nevertheless, Table~\ref{tab:three}
demonstrates that a measurement of the resonance curve with a
luminosity of 100~fb$^{-1}$ may well be able to eliminate a number of
bulk RS models. At a luminosity upgraded LHC it should be possible to
measure the couplings of a KK gluon candidate rather well, and, perhaps,
uniquely identify the underlying bulk RS model nature may have chosen.

\subsection{KK gluon coupling analysis}
\label{sec:threec}

The interactions of KK gluons in the $E_i$ models and models with a
large brane kinetic term $\kappa r_{IR}$ are characterized by
four couplings, $g^q$, $g_L^b=g_L^t$, $g_R^b$, and $g_R^t$. In all other
models considered here, $g_R^b=g^q$, and there are only three
independent couplings. A precise measurement of the Breit-Wigner
resonance curve of KK gluons should make it possible to determine at
least some of the couplings of KK gluons. Since the $b$-quark parton
densities are
much smaller than those of the light quarks, $b\bar b\to G\to t\bar t$
contributes little to the KK gluon cross section, even when $g_{L,R}^b$
is much larger than the SM strong coupling constant (eg. in the $E_2$
and $E_4$ models). This makes it essentially impossible to directly measure
$g^b_{L,R}$. However, $g^q$, $g_L^t$ and $g_R^t$ can, in principle, be 
measured. 

The dependence of the $t\bar t$ cross section on the KK gluon couplings
is of Breit-Wigner form. Since the width of the KK gluons depends
on the coupling constants, the dependence of the $t\bar t$ cross section
on the KK gluon couplings is sufficiently complicated to make the numerical
extraction of sensitivity bounds very CPU time consuming
when all three couplings are varied simultaneously. We do not attempt
such a general analysis here. Instead, in order to get a general idea of
how well the couplings of a spin~1 resonance in the $t\bar t$ channel
may be determined at the LHC, we derive
sensitivity limits for the following two limiting cases which are of
interest for the models discussed here, and which greatly simplify the
numerical analysis. 

\begin{enumerate}

\item The total width of the resonance (see Eq.~(\ref{eq:width})) is
dominated by one 
coupling. Models which fall into this category are the basic RS, $E_1$
and $E_3$ models where $g_R^t$ dominates the width, the $SO(5)$, $N=0$,
model where $g_L^t$ dominates, and the models with a large brane kinetic
term $\kappa r_{IR}$ where the width is dominated by $g^q$. Since the
contributions of the other two couplings to the total width is
negligible, the cross section is approximately bi-linear in these
couplings. This makes it possible to analytically solve for the
coefficients multiplying these couplings in each bin of the distributions
which are analyzed, provided that the coupling which dominates the width
is treated as a constant. These coefficients are valid for arbitrary
values of those 
couplings which are varied, even in regions where the dependence of the total
width on those couplings can no longer be neglected. As a result, 
it becomes straightforward to derive one- and
two-dimensional sensitivity bounds for these couplings. In
order to ensure that our results remain valid for large deviations of
the couplings from their predicted values, we do take into account the
dependence of the width on the couplings when deriving limits. Whenever we
derive bounds for those couplings which have a negligible impact on the
total width of the KK gluon, we assume that the third coupling (which
dominates the width) has the default value predicted by the model
considered. 

Naively, one may think that the cross section should be most sensitive
to the coupling which dominates the total width, $g_{dom}$. However, this
is not the case. Most of the sensitivity comes from the immediate
vicinity of the resonance, $m(t\bar t)=M_G$. At the resonance peak, the
dependence of the numerator and the denominator on $g_{dom}$ in the
square of the KK gluon amplitude approximately cancels. In addition, 
the interference term between the KK gluon and the SM amplitude vanishes
for $m(t\bar t)=M_G$. As a result,
the cross section is quite insensitive to the coupling which dominates
the total width. 

In the following, we will derive sensitivity limits for $g_{dom}$,
assuming that the two other couplings are fixed to the values 
characteristic for the model under consideration. 

\item In the remaining models, each coupling, unless it grossly deviates
from its predicted value, has only a small effect on the total width. In
this case we follow the approach outlined above for such couplings and
derive one- and two-dimensional sensitivity limits. 

\end{enumerate}

In the following we present 68.3\% confidence level (CL) limits for
$g^q$, $g_L^b=g_L^t$, and $g_R^t$, and $M_G=3$~TeV. A KK gluon with a
mass of 3~TeV can be discovered with a $5\sigma$ significance or better
in all models considered here, except the $SO(5)$ model with $N=1$. 
We derive 
limits for integrated luminosities of 100~fb$^{-1}$ and 300~fb$^{-1}$ at
the LHC, and 3000~fb$^{-1}$ at the SLHC. As before, we combine
information from the $m(t\bar t)$ and the $p_T(t\to b\ell\nu)$
distributions. For smaller (larger) KK gluon masses, more (less)
stringent limits on the couplings are obtained.

Sensitivity limits for the case when only one coupling at a time is
varied are presented in Tables~\ref{tab:four} and~\ref{tab:five}.
\begin{table}
\caption{$68.3\%$ CL limits for the couplings of a KK gluon with mass
$M_G=3$~TeV for various integrated luminosities at the LHC and
SLHC. Results are show for the basic RS model, the $SO(5)$ model with
$N=0$ and $N=1$, and two models with a large brane kinetic term $\kappa
r_{IR}$. Only one coupling at a time is varied. All limits are given in
units of the QCD coupling constant $g_s$. } 
\label{tab:four}
\vspace{2mm}
\begin{tabular}{c|ccc|c|ccc}
\multicolumn{4}{c|}{basic RS model} & \multicolumn{4}{c}{$SO(5)$} 
\\
\tableline
$\int{\cal L}dt$ & 100~fb$^{-1}$ & 300~fb$^{-1}$ & 3000~fb$^{-1}$ &
$\int{\cal L}dt$ & \phantom{$N=0$} 100~fb$^{-1}$ & 300~fb$^{-1}$ &
3000~fb$^{-1}$ \\ 
\tableline
$g^q=-0.2$ & $\begin{array}{c} +0.018 \\[-4pt] {-0.014}\end{array}$ &
$\begin{array}{c} +0.009 \\[-4pt] -0.010\end{array}$ &
$\begin{array}{c} +0.003 \\[-4pt] -0.003\end{array}$ & $g^q=-0.2$ &
$\begin{array}{c} N=0~\begin{array}{c} +0.026 \\[-4pt] -0.018
\end{array}\\ N=1~\begin{array}{c} +0.068 \\[-4pt] -0.032
\end{array}\end{array}$ &  
$\begin{array}{c} \begin{array}{c} +0.016 \\[-4pt]
-0.010 \end{array}\\ \begin{array}{c} +0.036 \\[-4pt]
-0.022 \end{array} \end{array}$ &
$\begin{array}{c} \begin{array}{c} +0.004 \\[-4pt]
-0.005 \end{array} \\ \begin{array}{c} +0.010 \\[-4pt]
-0.010 \end{array} \end{array} $ \\ 
$g_L^t=1$ & $\begin{array}{c} +1.07 \\[-4pt] {-0.50}\end{array}$ &
$\begin{array}{c} +0.66 \\[-4pt] -0.31\end{array}$ &
$\begin{array}{c} +0.18 \\[-4pt] -0.14\end{array}$ & $g_L^t=2.76$ &
$\begin{array}{c} N=0~\begin{array}{c} +0.65 \\[-4pt] -0.70
\end{array} \\ N=1~\begin{array}{c} +0.60 \\[-4pt] -0.52
\end{array} \end{array} $ & 
$\begin{array}{c} \begin{array}{c} +0.42 \\[-4pt] -0.35
\end{array} \\ \begin{array}{c} +0.41 \\[-4pt] -0.36
\end{array} \end{array} $ &
$\begin{array}{c} \begin{array}{c} +0.05 \\[-4pt] -0.04
\end{array} \\ \begin{array}{c} +0.23 \\[-4pt] -0.17
\end{array} \end{array} $ \\
$g_R^t=4$ & $\begin{array}{c} +0.64 \\[-4pt] {-0.80}\end{array}$ &
$\begin{array}{c} +0.21 \\[-4pt] -0.39\end{array}$ &
$\begin{array}{c} +0.05 \\[-4pt] -0.04\end{array}$ & $g_R^t=0.07$ & 
$\begin{array}{c} N=0~\begin{array}{c} +0.37 \\[-4pt] -0.34
\end{array} \\ N=1~\begin{array}{c} +0.67 \\[-4pt] -0.47
\end{array} \end{array} $ & 
$\begin{array}{c} \begin{array}{c} +0.26 \\[-4pt] -0.22
\end{array} \\ \begin{array}{c} +0.51 \\[-4pt] -0.32
\end{array} \end{array} $ &
$\begin{array}{c} \begin{array}{c} +0.14 \\[-4pt] -0.12
\end{array} \\ \begin{array}{c} +0.24 \\[-4pt] -0.16
\end{array} \end{array} $ \\
\tableline
\multicolumn{4}{c|}{$\kappa r_{IR}=5$} & \multicolumn{4}{c}{$\kappa
r_{IR}=20$}\\
\tableline
$\int{\cal L}dt$ & 100~fb$^{-1}$ & 300~fb$^{-1}$ & 3000~fb$^{-1}$ &
$\int{\cal L}dt$ & 100~fb$^{-1}$ & 300~fb$^{-1}$ & 3000~fb$^{-1}$ \\
\tableline
$g^q=-0.4$ & $\begin{array}{c} +0.14 \\[-4pt] -0.09\end{array}$ &
$\begin{array}{c} +0.10 \\[-4pt] -0.06\end{array}$ &
$\begin{array}{c} +0.02 \\[-4pt] -0.03\end{array}$ & $g^q=-0.8$ & 
$\begin{array}{c} +0.36 \\[-4pt] -0.15\end{array}$ &
$\begin{array}{c} +0.19 \\[-4pt] -0.09\end{array}$ &
$\begin{array}{c} +0.03 \\[-4pt] -0.02\end{array}$ \\
$g_L^t=-0.2$ & $\begin{array}{c} +0.38 \\[-4pt] -0.11\end{array}$ &
$\begin{array}{c} +0.28 \\[-4pt] -0.07\end{array}$ &
$\begin{array}{c} +0.05 \\[-4pt] -0.04\end{array}$ & $g_L^t=-0.6$ &
$\begin{array}{c} +0.05 \\[-4pt] -0.06\end{array}$ &
$\begin{array}{c} +0.04 \\[-4pt] -0.02\end{array}$ &
$\begin{array}{c} +0.01 \\[-4pt] -0.01\end{array}$ \\
$g_R^t=0.6$ & $\begin{array}{c} +0.05 \\[-4pt] -0.07\end{array}$ &
$\begin{array}{c} +0.03 \\[-4pt] -0.04\end{array}$ &
$\begin{array}{c} +0.01 \\[-4pt] -0.01\end{array}$ & $g_R^t=-0.2$ &
$\begin{array}{c} +0.21 \\[-4pt] -0.14\end{array}$ &
$\begin{array}{c} +0.13 \\[-4pt] -0.06\end{array}$ &
$\begin{array}{c} +0.03 \\[-4pt] -0.03\end{array}$ \\
\end{tabular}
\end{table}
\begin{table}[h!]
\caption{$68.3\%$ CL limits for the couplings of a KK gluon with mass
$M_G=3$~TeV for various integrated luminosities at the LHC and
SLHC. Results are show for the $E_i$, $i=1,\dots,4$ models. Only one
coupling at a time is varied. All limits are given in 
units of the QCD coupling constant $g_s$. $E_1$ and $E_2$ ($E_3$ and
$E_4$) KK gluons differ only in the strength of their coupling to
right-handed $b$-quarks, see Table~\ref{tab:one}.} 
\label{tab:five}
\vspace{2mm}
\begin{tabular}{c|ccc|c|ccc}
\multicolumn{4}{c|}{$E_1$} & \multicolumn{4}{c}{$E_2$}\\
\tableline
$\int{\cal L}dt$ & 100~fb$^{-1}$ & 300~fb$^{-1}$ & 3000~fb$^{-1}$ &
$\int{\cal L}dt$ & 100~fb$^{-1}$ & 300~fb$^{-1}$ & 3000~fb$^{-1}$ \\
\tableline
$g^q=-0.2$ & $\begin{array}{c} +0.018 \\[-4pt] -0.012\end{array}$ &
$\begin{array}{c} +0.010 \\[-4pt] -0.008\end{array}$ &
$\begin{array}{c} +0.003 \\[-4pt] -0.003\end{array}$ & $g^q=-0.2$ &
$\begin{array}{c} +0.023 \\[-4pt] -0.015\end{array}$ &
$\begin{array}{c} +0.012 \\[-4pt] -0.010\end{array}$ &
$\begin{array}{c} +0.004 \\[-4pt] -0.004\end{array}$ \\
$g_L^t=1.34$ & $\begin{array}{c} +0.72 \\[-4pt] -0.44\end{array}$ &
$\begin{array}{c} +0.58 \\[-4pt] -0.32\end{array}$ &
$\begin{array}{c} +0.23 \\[-4pt] -0.14\end{array}$ & $g_L^t=1.34$ &
$\begin{array}{c} +0.90 \\[-4pt] -0.83\end{array}$ &
$\begin{array}{c} +0.65 \\[-4pt] -0.57\end{array}$ &
$\begin{array}{c} +0.24 \\[-4pt] -0.22\end{array}$ \\
$g_R^t=4.9$ & $\begin{array}{c} +0.90 \\[-4pt] -0.90\end{array}$ &
$\begin{array}{c} +0.53 \\[-4pt] -0.42\end{array}$ &
$\begin{array}{c} +0.22 \\[-4pt] -0.14\end{array}$ & $g_R^t=4.9$ &
$\begin{array}{c} +0.80 \\[-4pt] -1.02\end{array}$ &
$\begin{array}{c} +0.58 \\[-4pt] -0.64\end{array}$ &
$\begin{array}{c} +0.24 \\[-4pt] -0.21\end{array}$ \\
\tableline
\multicolumn{4}{c|}{$E_3$} & \multicolumn{4}{c}{$E_4$}\\
\tableline
$\int{\cal L}dt$ & 100~fb$^{-1}$ & 300~fb$^{-1}$ & 3000~fb$^{-1}$ &
$\int{\cal L}dt$ & 100~fb$^{-1}$ & 300~fb$^{-1}$ & 3000~fb$^{-1}$ \\
\tableline 
$g^q=-0.2$ & $\begin{array}{c} +0.020 \\[-4pt] -0.014\end{array}$ & 
$\begin{array}{c} +0.011 \\[-4pt] -0.008\end{array}$ & 
$\begin{array}{c} +0.003 \\[-4pt] -0.003\end{array}$ & $g^q=-0.2$ & 
$\begin{array}{c} +0.026 \\[-4pt] -0.017\end{array}$ & 
$\begin{array}{c} +0.014 \\[-4pt] -0.008\end{array}$ & 
$\begin{array}{c} +0.004 \\[-4pt] -0.004\end{array}$ \\ 
$g_L^t=1.34$ & $\begin{array}{c} +0.54 \\[-4pt] -0.64\end{array}$ &
$\begin{array}{c} +0.35 \\[-4pt] -0.38\end{array}$ &
$\begin{array}{c} +0.11 \\[-4pt] -0.12\end{array}$ & $g_L^t=1.34$ &
$\begin{array}{c} +0.76 \\[-4pt] -0.74\end{array}$ &
$\begin{array}{c} +0.48 \\[-4pt] -0.47\end{array}$ &
$\begin{array}{c} +0.18 \\[-4pt] -0.16\end{array}$ \\
$g_R^t=3.25$ & $\begin{array}{c} +0.66 \\[-4pt] -0.80\end{array}$ &
$\begin{array}{c} +0.37 \\[-4pt] -0.43\end{array}$ &
$\begin{array}{c} +0.12 \\[-4pt] -0.12\end{array}$ & $g_R^t=3.25$ &
$\begin{array}{c} +0.50 \\[-4pt] -0.68\end{array}$ &
$\begin{array}{c} +0.32 \\[-4pt] -0.52\end{array}$ &
$\begin{array}{c} +0.14 \\[-4pt] -0.16\end{array}$ \\
\end{tabular}
\end{table}
In all models, except those with a large brane kinetic term $\kappa
r_{IR}$, the coupling to light quarks can be measured with a precision
of $10-15\%$ for 100~fb$^{-1}$, and to 5\% or better for 3000~fb$^{-1}$. In 
models with  a large brane kinetic term $\kappa r_{IR}$, decays into
light quarks dominate the width (see Table~\ref{tab:one}), and $g^q$ can
only be determined with an accuracy of $35-45\%$ ($4-8\%$) for
100~fb$^{-1}$ (3000~fb$^{-1}$). Note that, in addition to the allowed
range for $g^q$ listed in Tables~\ref{tab:four} and~\ref{tab:five}, an
interval around $g^q=0$ cannot be excluded. 

Similarly, the coupling to left-handed top quarks can be measured with
a precision of $10-100\%$ ($2-20\%$) for 100~fb$^{-1}$ (3000~fb$^{-1}$)
except in the model with $\kappa r_{IR}=5$ where more than 300~fb$^{-1}$
are needed in order to rule out a vanishing of $g_L^t$. Similar
accuracies are achievable for $g_R^t$, except in the $SO(5)$ model where
$g_R^t$ almost vanishes and it will be impossible to establish a
non-vanishing coupling of the KK gluons to right-handed top quarks even
at the SLHC. 

The bounds on $g^t_{L,R}$ are in many cases significantly weaker than
those for $g^q$. In many of the models considered here, $g^t_L\ll g^t_R$
with $g^t_R$ being the coupling which dominates the width, or $g^t_R\ll
g^t_L$, and $g^t_L$ dominates the KK gluon width. Since the differential
cross section contains terms proportional to $g^{t2}_L+g^{t2}_R$ and
$g^t_L+g^t_R$, it is obvious that the sensitivity to $g^t_{L,R}$ is
significantly reduced in such models. 

\begin{figure}[th!] 
\begin{center}
\begin{tabular}{cc}
\includegraphics[width=8.2cm]{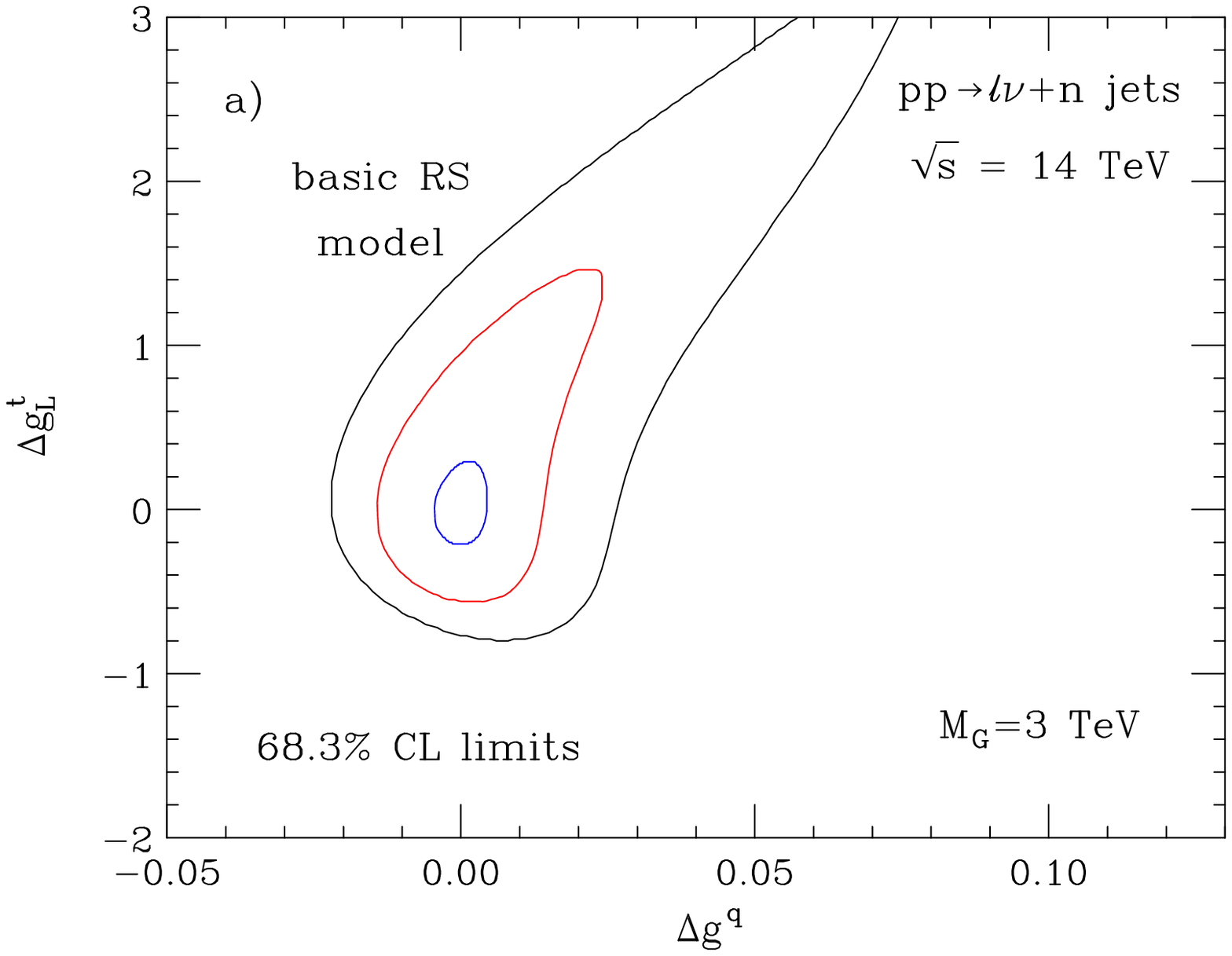} & 
\includegraphics[width=8.2cm]{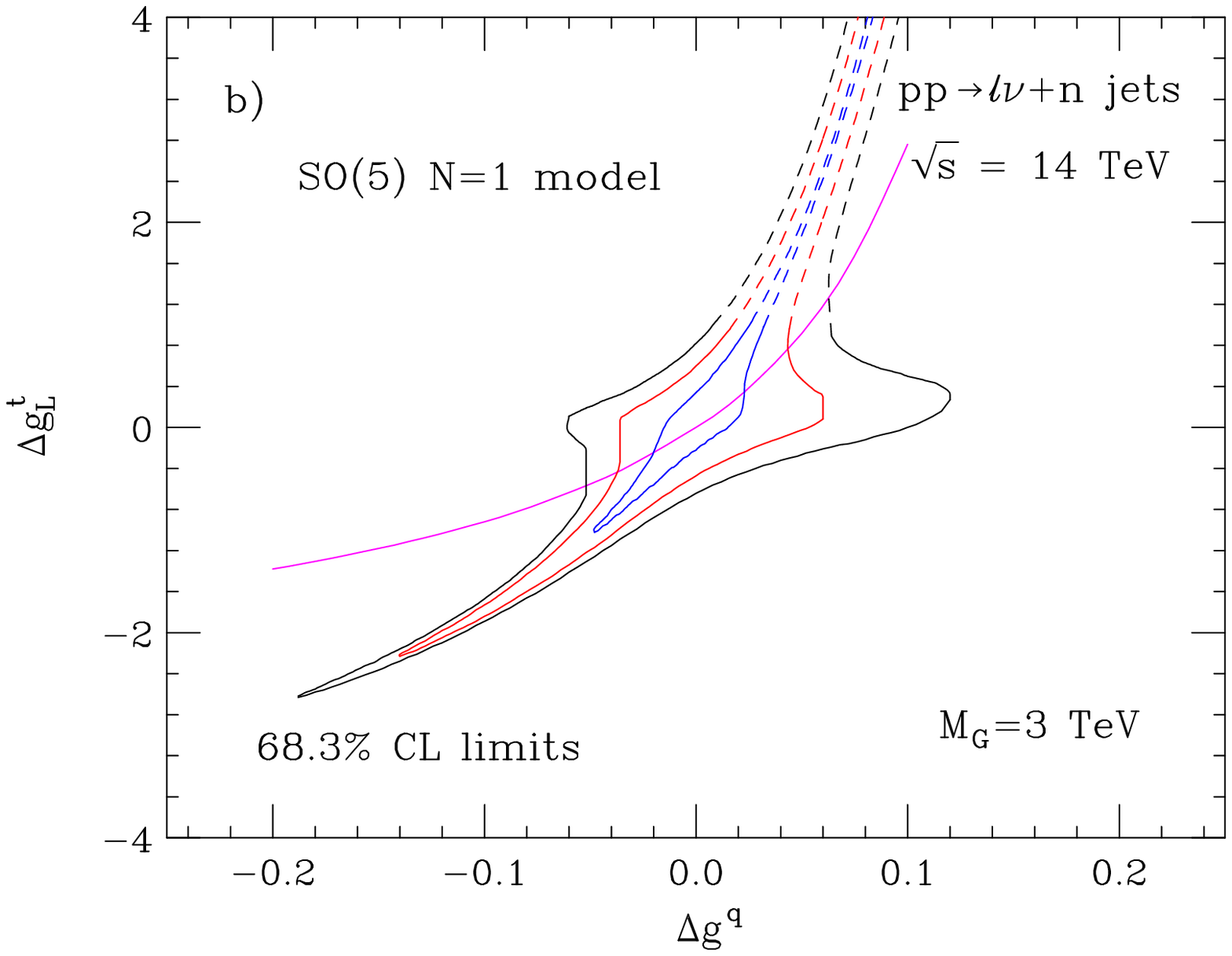} \\[4mm]
\includegraphics[width=8.2cm]{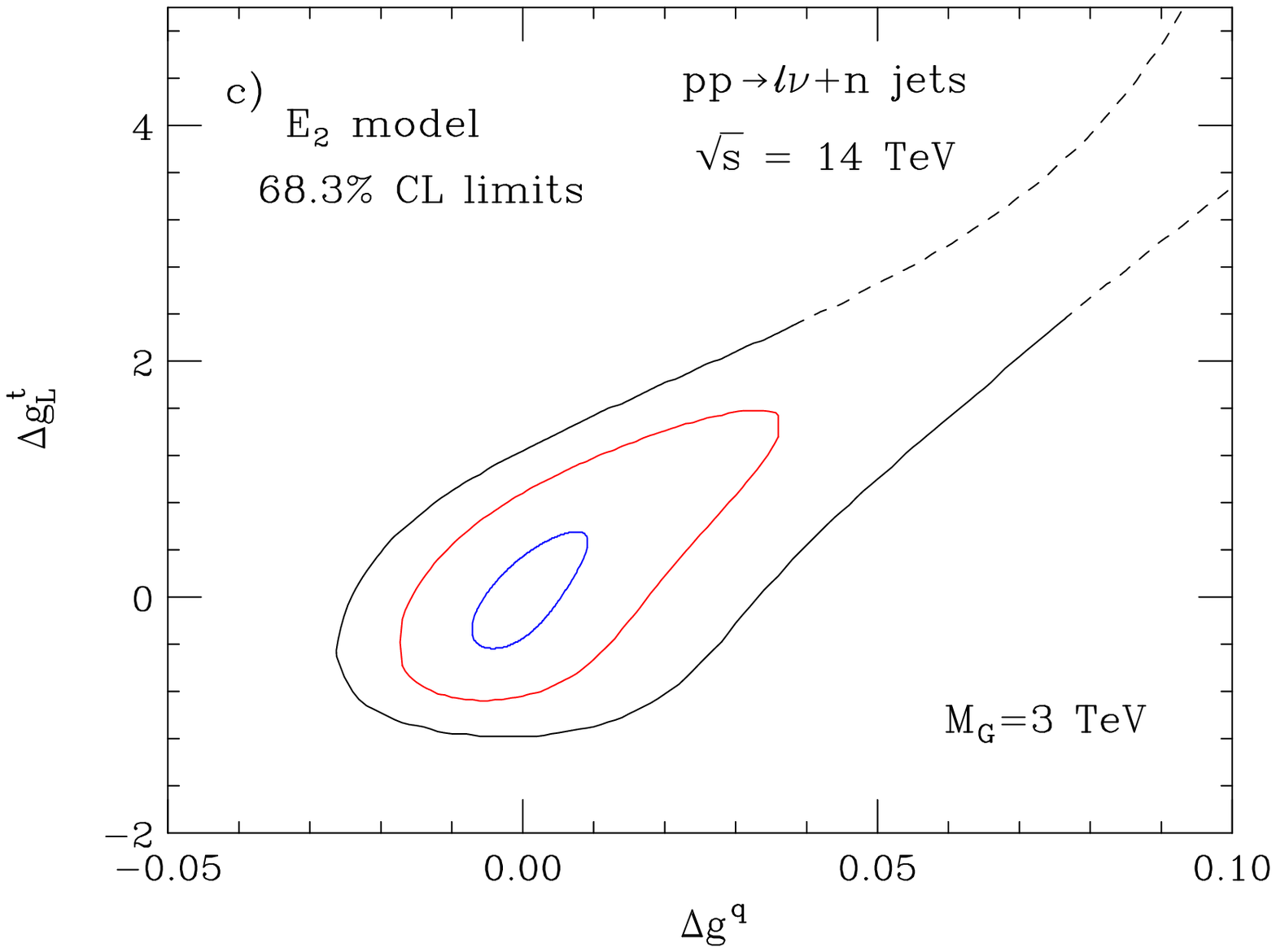} & 
\includegraphics[width=8.2cm]{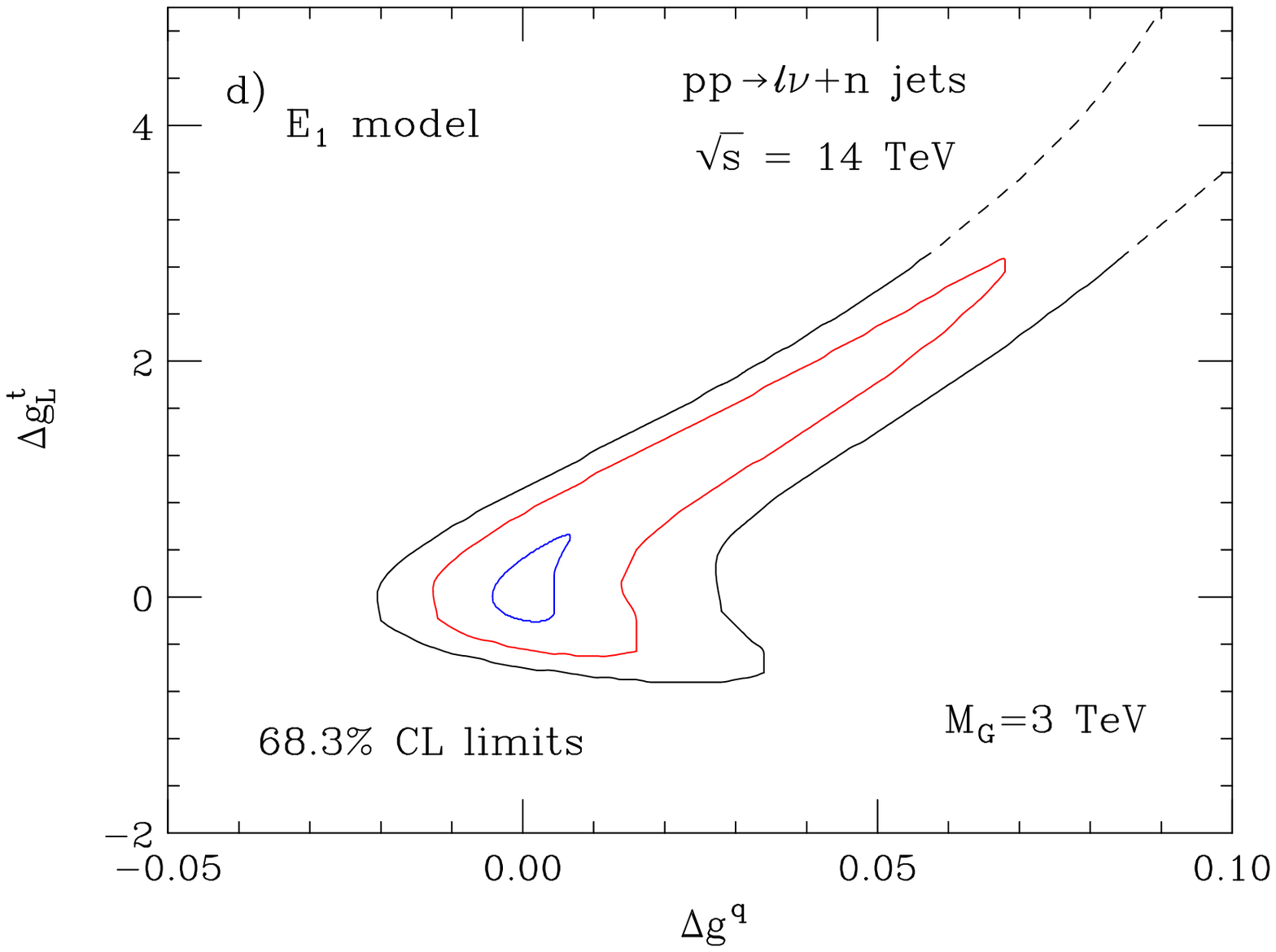} 
\end{tabular}
\vspace*{2mm}
\caption[]{\label{fig:three} 
Projected $68.3\%$ CL bounds on the couplings of KK gluons with
$M_G=3$~TeV to light
quarks, $g^q$, and left-handed top quarks, $g^t_L$, in a) the basic RS,
b) the $SO(5)$ with $N=1$, c) the $E_2$ and d) the $E_1$ model at the
LHC with an integrated luminosity of 100~fb$^{-1}$ (black lines),
300~fb$^{-1}$ (red lines), and 3000~fb$^{-1}$ (blue lines). The coupling
of the KK gluon to right-handed top quarks is assumed to have the
default value of the model considered (see Table~\ref{tab:one}). The
bounds are obtained from a 
log-likelihood analysis which combines information from the $m(t\bar t)$
and $p_T(t\to b\ell\nu)$ distributions. $\Delta g^q$ and $\Delta g^t_L$
are the deviations from the default values of the coupling constants
predicted by the model considered. The magenta line in part b) indicates
those couplings for which the product $g^qg^t_L$ is equal to the value
predicted for the $SO(5)$ model with $N=1$.
} 
\vspace{-7mm}
\end{center}
\end{figure}

By varying only one coupling at a time, we ignore correlations between different
couplings. These correlations are expected to be
particularly pronounced between $g^q$ and $g_{L,R}^t$. This is easy to
understand: the shape of the resonance curve may not change appreciably
if the magnitude of $g^q$ decreases, and that of the top 
quark coupling increases by a corresponding amount. Examples of
two-dimensional sensitivity limits in the $g^q-g_L^t$ plane are shown in 
Fig.~\ref{fig:three} for $M_G=3$~TeV. 

As expected, strong correlations are observed between $g^q$ and
$g^t_L$. In some cases, the limits weaken so much when both couplings
are varied simultaneously that $\Gamma_G/M_G$ becomes of ${\cal O}(1)$,
and one has to worry about $S$-matrix unitarity being violated. In this
region, the bounds on $g^q$ and $g^t_L$, of course, become
unreliable. The region in which $\Gamma_G/M_G>0.5$ is indicated by
dashed lines in Fig.~\ref{fig:three}. Note that the correlations between
the couplings become progressively smaller with increasing integrated
luminosity. 

The results shown for the $SO(5)$ model with $N=1$ deserve further
discussion. Figure~\ref{fig:three}b shows that it will be impossible to
place an upper bound on $g^t_L$. Even with 3000~fb$^{-1}$, a very narrow
funnel remains where it is not possible to distinguish 
$g^t_L$ and $g^q$ from the $SO(5)$ model with $N=1$. However, much of
that funnel lies in the region where possible unitarity violations cast
doubt on the reliability of our results. The peculiar shape of the
contour limits in the $SO(5)$ model with $N=1$ can be easily understood
by recalling that the coupling of the KK gluon to right-handed top
quarks almost vanishes in this model (see Table~\ref{tab:one}). In the
limit where $g_R^t=0$ and $\Gamma_G$ does not change appreciably
when $g^q$ and $g_L^t$ are varied, the Breit-Wigner resonance curve does
not change as long as the product $g^qg^t_L$ remains invariant. The
line of constant $g^qg^t_L$ is indicated by the magenta line in
Fig.~\ref{fig:three}b. In practice, the small but non-zero value of
$g^t_R=0.07$, and the variation of $\Gamma_G$ are responsible for the
deviation of the allowed coupling parameters from the line of constant
$g^qg^t_L$. The extremely strong correlations between $g^q$ and $g_L^t$
make it very difficult to pin down these couplings in the $SO(5)$ model
with $N=1$. Correlations between $g^q$ and $g_R^t$, and $g_L^t$ and
$g_R^t$, however, are small in this model.

\begin{figure}[th!] 
\begin{center}
\includegraphics[width=12.2cm]{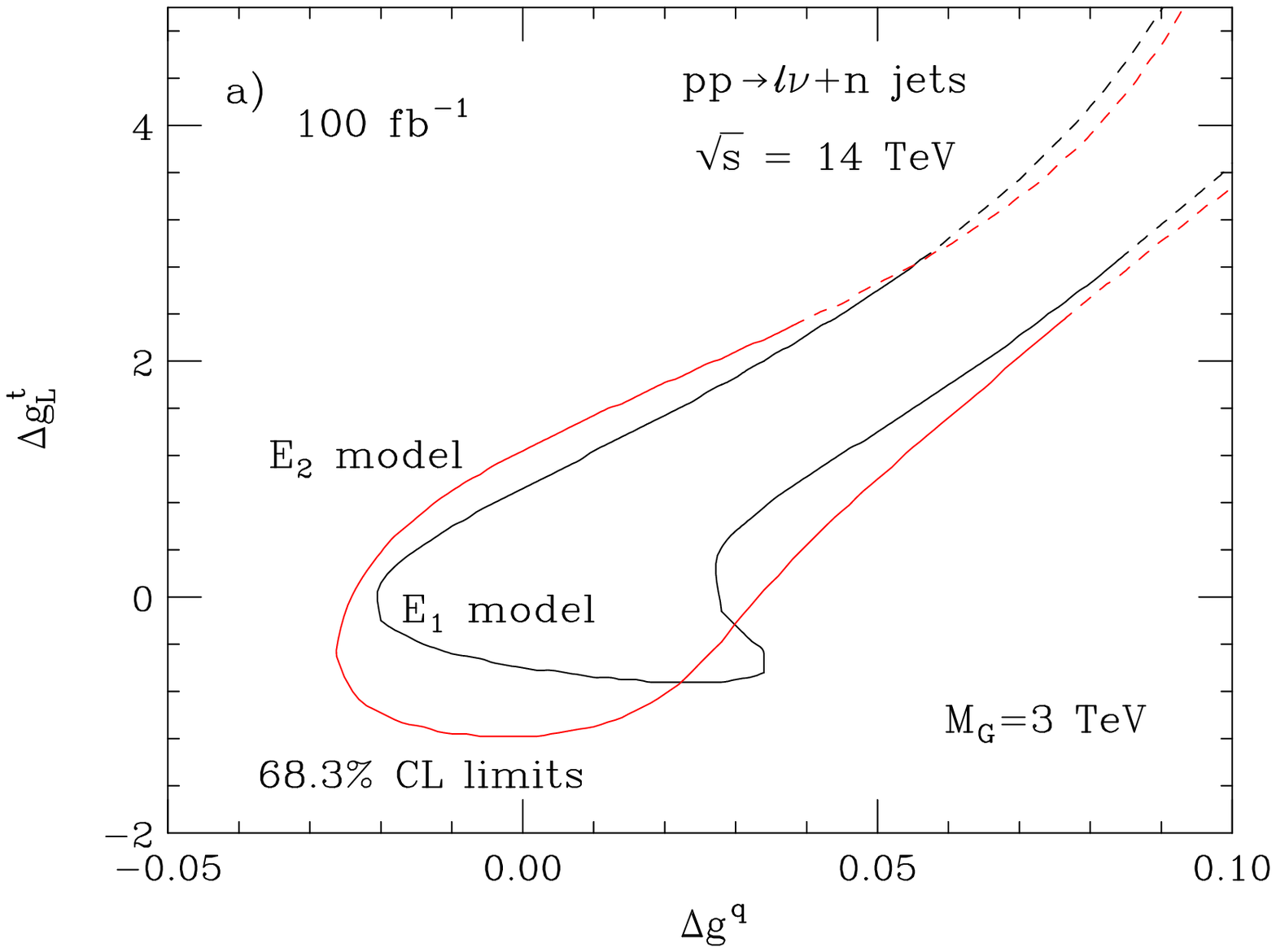} \\[3mm]
\includegraphics[width=12.2cm]{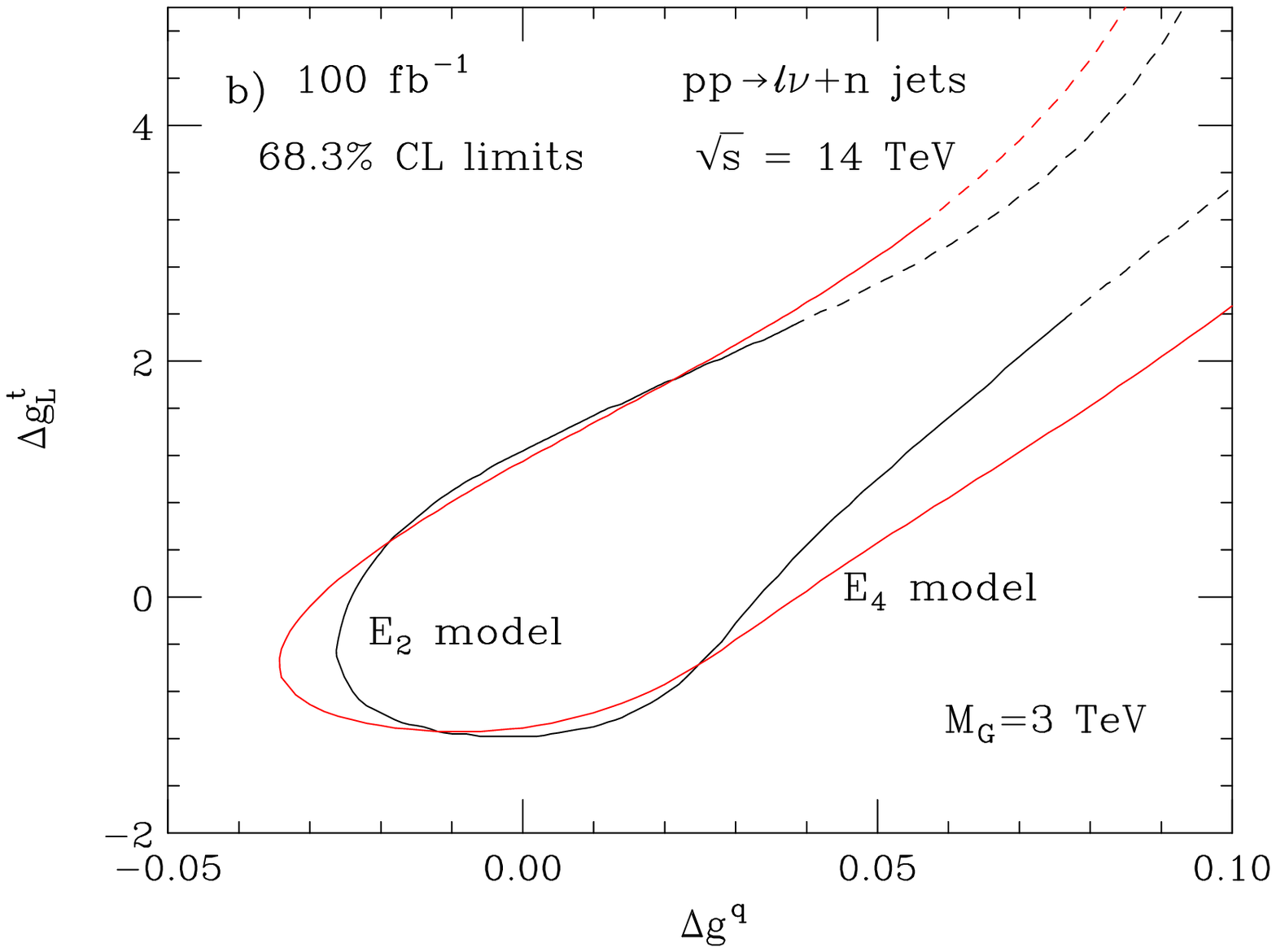}
\vspace*{2mm}
\caption[]{\label{fig:four} 
Projected $68.3\%$ CL bounds at the LHC on the couplings of KK gluons to light
quarks, $g^q$, and left-handed top quarks, $g^t_L$, in a) the $E_1$ and
$E_2$ models, and b) the $E_2$ and $E_4$ models. Results are shown for
$M_G=3$~TeV and 100~fb$^{-1}$. The coupling
of the KK gluon to right-handed top quarks is assumed to have the
default value of the respective model (see Table~\ref{tab:one}). The
bounds are obtained from a 
log-likelihood analysis which combines information from the $m(t\bar t)$
and $p_T(t\to b\ell\nu)$ distributions. $\Delta g^q$ and $\Delta g^t_L$
are the deviations from the default values of the coupling constants
predicted by the model considered. 
} 
\vspace{-7mm}
\end{center}
\end{figure}

In Fig.~\ref{fig:four} we compare the limits which can be achieved for
$g^q$ and $g^t_L$ with 100~fb$^{-1}$ in the $E_1$ and $E_2$ models, and
the $E_2$ and $E_4$ models, respectively. 
KK gluons in the $E_1$ and $E_2$ models differ only by their coupling to the
right-handed $b$-quarks, and the total width. Similarly, in the
$E_2$ and $E_4$ models, only the coupling of the KK gluons to right-handed
top quarks and the total width differ. Figure~\ref{fig:four} 
demonstrates that the sensitivity limits for $g^q$ and $g_L^t$
in the $E_i$ models depend only modestly on other coupling
parameters. Qualitatively similar results are obtained for 300~fb$^{-1}$
and 3000~fb$^{-1}$.

\begin{figure}[th!] 
\begin{center}
\includegraphics[width=11.7cm]{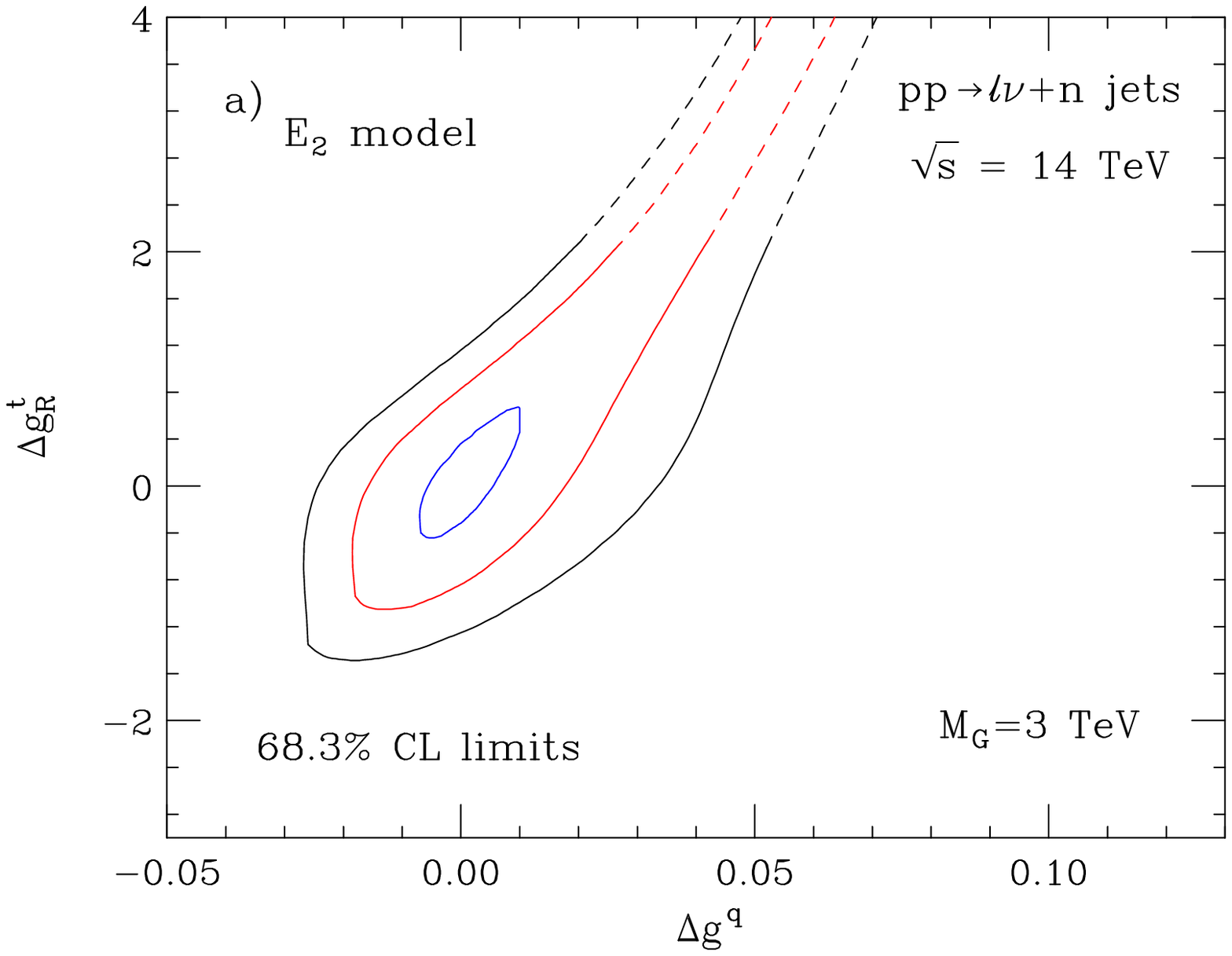} \\[3mm]
\includegraphics[width=11.7cm]{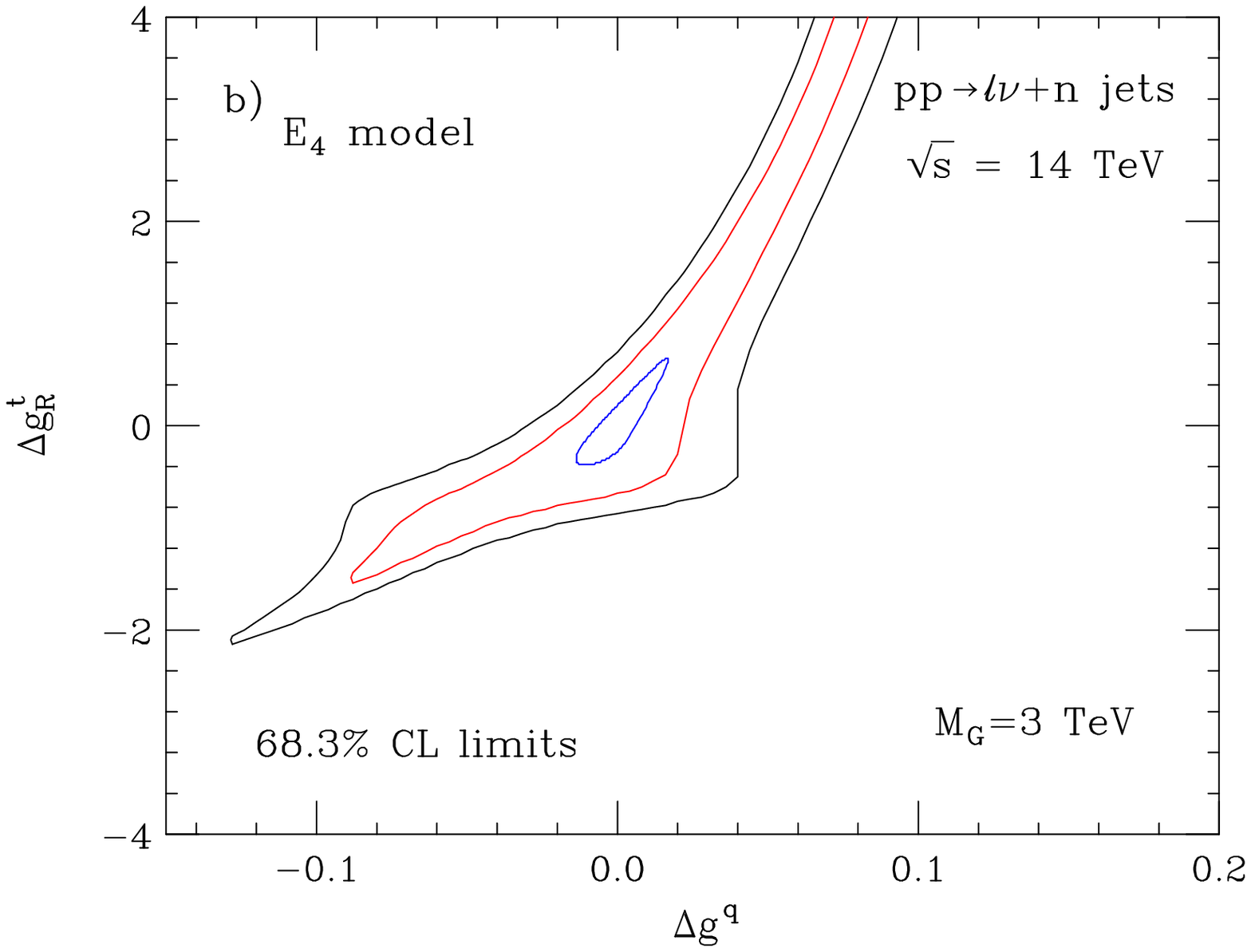}
\vspace*{2mm}
\caption[]{\label{fig:five} 
Projected $68.3\%$ CL bounds at the LHC on the couplings of KK gluons to light
quarks, $g^q$, and right-handed top quarks, $g^t_R$, in a) the $E_2$,
and b) the $E_4$ model, with an integrated luminosity of 100~fb$^{-1}$
(black lines), 
300~fb$^{-1}$ (red lines), and 3000~fb$^{-1}$ (blue lines). The coupling
of the KK gluon to left-handed top quarks is assumed to have the
default value of the model considered (see Table~\ref{tab:one}). The
bounds are obtained from a 
log-likelihood analysis which combines information from the $m(t\bar t)$
and $p_T(t\to b\ell\nu)$ distributions. $\Delta g^q$ and $\Delta g^t_R$
are the deviations from the default values of the coupling constants
predicted by the model considered. 
} 
\vspace{-7mm}
\end{center}
\end{figure}

Strong correlations may also occur between $g^q$ and $g_R^t$. As an
example, we show the two dimensional 68.3\% CL sensitivity
limits in the $g^q-g_R^t$ plane for the $E_2$ and $E_4$ models and
$M_G=3$~TeV in Fig.~\ref{fig:five}.
In order to pin down $g_R^t$ with a precision of ${\cal O}(10\%)$ in these
models, a luminosity upgrade of the LHC is needed. Similar correlations are
observed between $g^q$ and $g_L^t$ in the two models (see
Fig.~\ref{fig:four}). On the other hand, $g_L^t$ and $g_R^t$ display little
correlation. 

\begin{figure}[th!] 
\begin{center}
\includegraphics[width=11.6cm]{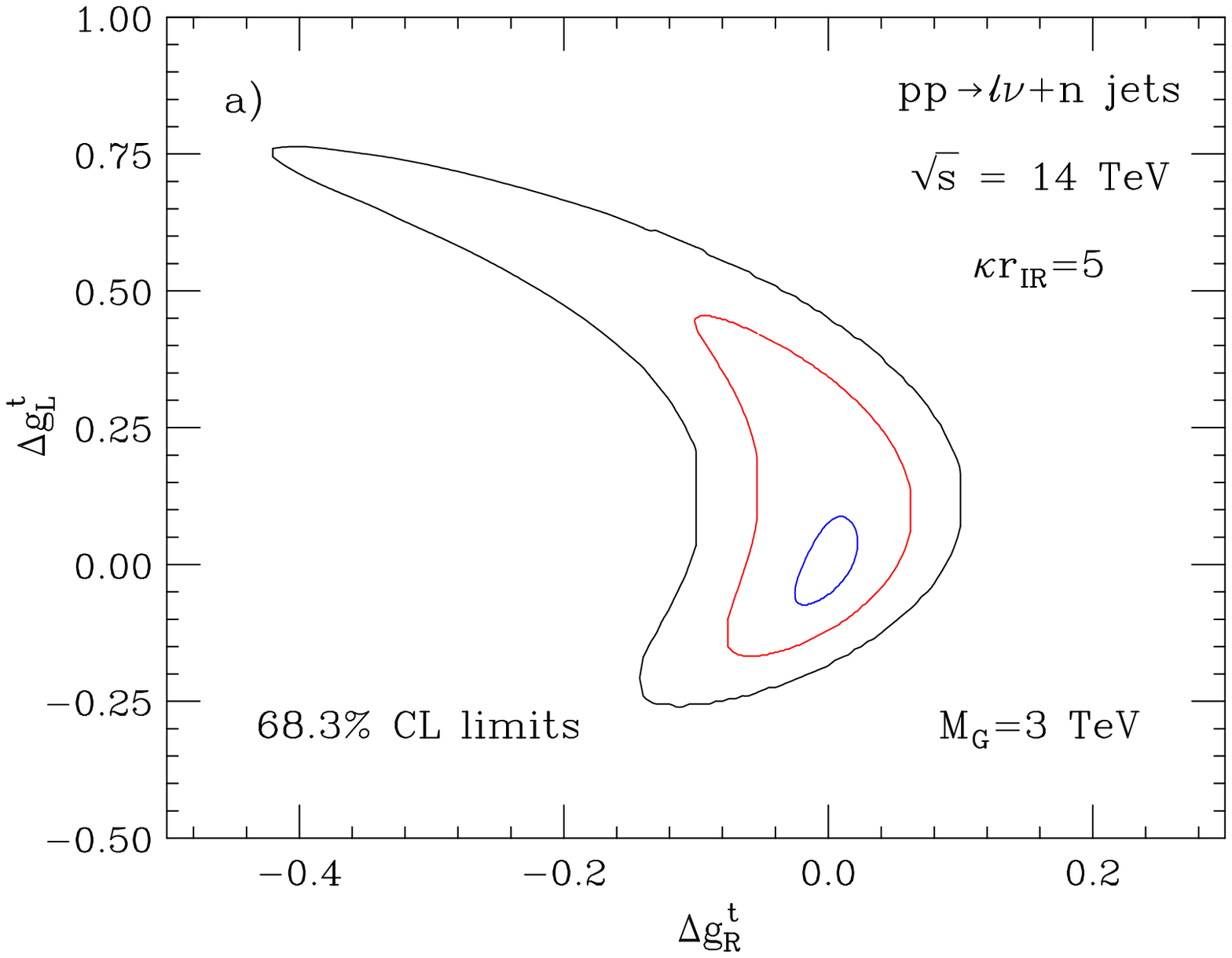} \\[3mm]
\includegraphics[width=11.6cm]{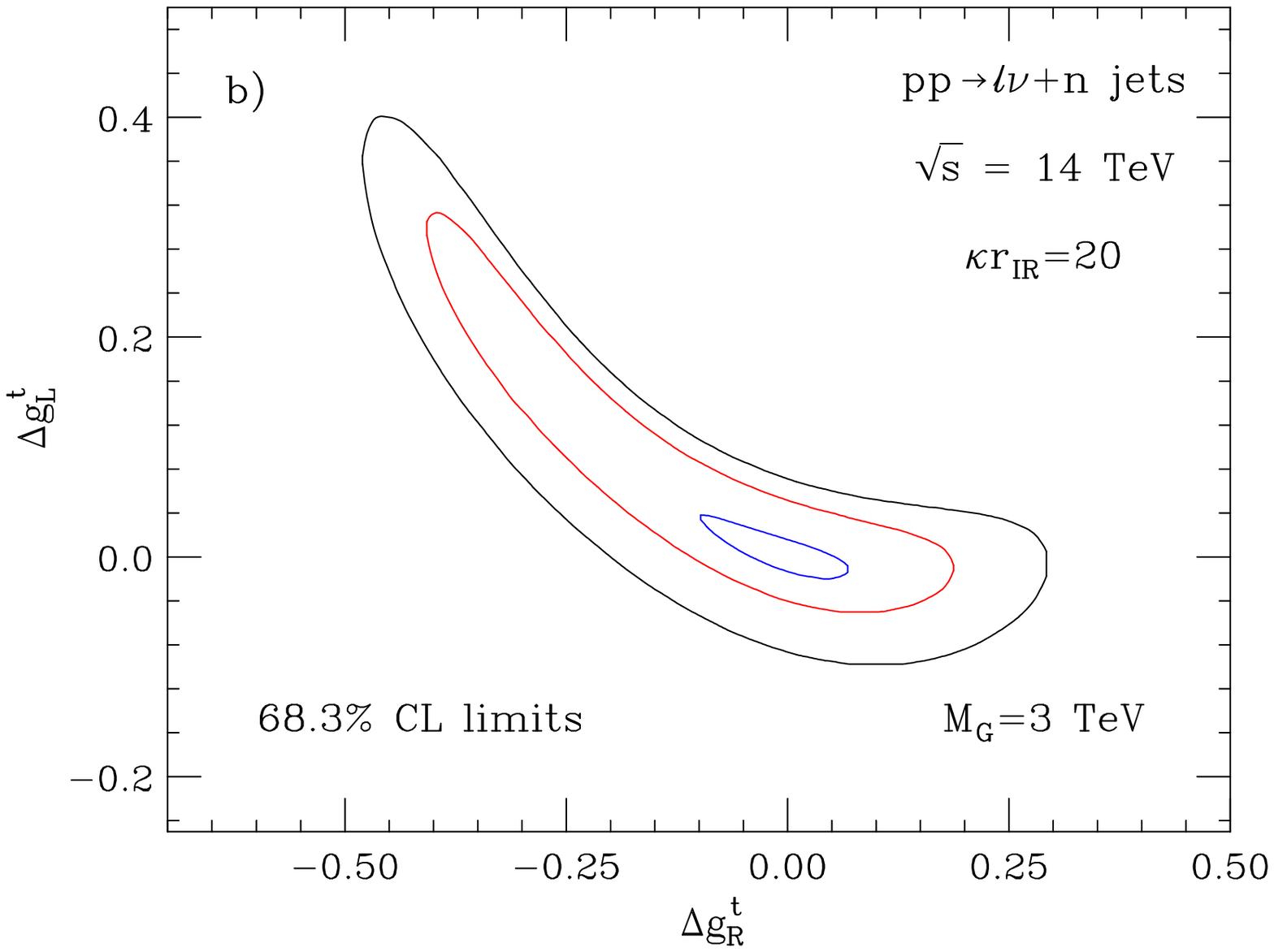}
\vspace*{2mm}
\caption[]{\label{fig:six} 
Projected $68.3\%$ CL bounds at the LHC on the couplings of KK gluons to
left- and right-handed top quarks, $g^t_L$ and $g^t_R$ in two models
with a large brane kinetic term $\kappa r_{IR}$. Results are shown for
100~fb$^{-1}$ (black lines), 
300~fb$^{-1}$ (red lines), and 3000~fb$^{-1}$ (blue lines). The mass of
the KK gluon is fixed to $M_G=3$~TeV. The coupling
of the KK gluon to light quarks is assumed to have the
default value of the model considered (see Table~\ref{tab:one}). The
bounds are obtained from a 
log-likelihood analysis which combines information from the $m(t\bar t)$
and $p_T(t\to b\ell\nu)$ distributions. $\Delta g^t_L$ and $\Delta g^t_R$
are the deviations from the default values of the coupling constants
predicted by the model considered. 
} 
\vspace{-7mm}
\end{center}
\end{figure}

However, strong correlations between couplings are not only observed
between $g^q$ and $g^t_{L,R}$, but also between the couplings of KK
gluons to left- and right-handed top quarks. Figure~\ref{fig:six} shows
68.3\% CL limits for $g^t_L$ and $g^t_R$ in two models with a large brane
kinetic term $\kappa r_{IR}$. 

Figures~\ref{fig:three} --~\ref{fig:six} demonstrate that
one-dimensional limits on the couplings of KK gluons may be totally
misleading. Although the correlations between couplings become
progressively smaller with increasing integrated luminosity, they may still
significantly weaken sensitivity limits at the SLHC, in some cases by up
to a factor~3. Although we have not studied the correlations for cases
where one of the couplings dominates the KK gluon width, we expect that
strong correlations may also occur there. While it will only be possible to
obtain a limited amount of information on the couplings of KK gluons at
the LHC with 300~fb$^{-1}$ or less of data when correlations are
included, it will be possible to measure them with a precision of
$5-50\%$ at the SLHC. 

\section{Discussion and Conclusions}
\label{sec:four}

Many New Physics models predict the existence of new particles decaying
into a $t\bar t$ pair
with masses in the TeV region. They lead to a peak in the $t\bar t$
invariant mass distribution and a Jacobian peak in the $p_T(t)$
differential cross section. In this paper we specifically studied the
production of KK gluons in bulk RS models in the $t\bar t$ channel at
the LHC. Since the couplings of KK gluons to light quarks is suppressed
in many bulk RS models, the $t\bar t$ final state becomes their main
discovery channel. The lepton+jets final state offers a good opportunity
to search for such particles. 

The search for resonances in the $t\bar t$ channel with masses in the
TeV region requires the reconstruction of very energetic top quarks
which faces two major difficulties. Firstly, very energetic top quarks
are strongly
boosted, and their decay products are highly collimated. This leads to
overlapping and merging jets from hadronically decaying top
quarks. Secondly, the tagging efficiency
for $b$-quarks in $t\bar t$ events with very energetic top quarks may be
up to a factor~3 smaller, and the misidentification probability of light
quark or gluon jets may be up to a factor of~3 higher, than at low
energies. This reduces the number of $t\bar t$ events which can be
identified, and increases the background. 

As we have shown in Ref.~\cite{Baur:2007ck}, these problems can be
partially overcome by considering the $\ell\nu+n$~jets final states with
one or two tagged $b$-quarks and $n=2,\,3,\,4$ instead of the canonical
$\ell\nu+4$~jets final state with two $b$-tags, and by imposing
suitable invariant mass and cluster transverse mass cuts. Using the
results of Ref.~\cite{Baur:2007ck}, we calculated $5\sigma$ discovery
limits for KK gluons in nine different bulk RS models by combining
information from the $t\bar t$ invariant mass, and the $p_T(t\to
b\ell\nu)$ distribution. Although information on the longitudinal degree
of freedom is lost in the $p_T(t\to b\ell\nu)$ distribution, it has the
advantage of a substantially smaller SM non-$t\bar t$ background. Our
calculation takes into account the typical momentum resolution of an LHC
experiment, particle identification efficiencies, and the energy loss
due to $b$-quark decay. 

Assuming a $b$-tagging efficiency of $\epsilon_b=0.2$ and a light
quark/gluon jet misidentification probability of $P_{j\to b}=1/30$, as
suggested by preliminary ATLAS simulations~\cite{atlastdr,cmstdr}, we
found that, in most models considered, KK gluons with a mass of up to
$3.5-4$~TeV ($4-4.5$~TeV) can be discovered at the LHC with an
integrated luminosity of 100~fb$^{-1}$ (300~fb$^{-1}$). For comparison,
electroweak precision measurements require KK gluons in bulk RS models
to be heavier than $2-3$~TeV~\cite{Carena:2006bn}. The LHC should
therefore be able to considerably constrain such models.

For comparison, we also listed the discovery limits for the $Z_H$ boson in
the Littlest Higgs model, and the KK graviton in bulk RS models in the
$t\bar t$ channel. The discovery limits for the $Z_H$ boson are about
a factor~1.5, and those for the KK graviton are more than a factor~2,
weaker than those for KK gluons. In both cases, other final states may offer
a better chance to search for these particles: the $Z_H$ boson can be
discovered in Drell-Yan production with masses up to 5~TeV, whereas a
KK graviton in bulk RS models can be found in the $WW$ final state with
masses up to 3.5~TeV. 

We also investigated, for the example of a KK gluon with mass
$M_G=3$~TeV, how well different bulk RS models can be 
distinguished through a measurement of the KK gluon resonance curve. We
found that, for 100~fb$^{-1}$,  the 
$E_i$ models can only be distinguished at the $1.5-3\sigma$
level. However, the remaining models can be discriminated with a
significance of $4-10\sigma$. The $E_i$ models and non-$E_i$ models,
finally, can be separated at the $2-11\sigma$ level, except for the
basic RS and the $E_3$ model which will be very hard to discriminate from a
measurement of the KK gluon resonance curve. The conclusion to draw from
this investigation is that the Breit-Wigner resonance curve in the
$t\bar t$ final state does have some analyzing power, and thus 
may be helpful in discriminating new physics models. 

Finally, we studied how well the KK gluon couplings can be measured at
the LHC and SLHC. In the $E_i$, $i=1,\dots, 4$ models, the coupling to
the right-handed $b$-quark is an independent parameter. Since $b$-quark
fusion contributes only little to the KK gluon cross section, it will be
impossible to determine the $Gb_Rb_R$ coupling from the shape of the KK gluon
resonance curve in the $t\bar t$ final state. The remaining three
couplings, $g^q$, $g^b_L=g_L^t$, and $g^t_R$, however, can be
constrained from an analysis of the $m(t\bar t)$ and $p_T(t\to
b\ell\nu)$ distributions. We presented one- and two-dimensional 68.3\%
CL limits for these couplings. In several models, one coupling
completely dominates the KK gluon width. Since interference effects
vanish, and the dependence on the coupling which dominates the width
approximately cancels, at the peak position of the Breit-Wigner resonance
where most KK gluon events are concentrated, it will be difficult to
precisely measure this coupling. We also found that correlations between
couplings may strongly affect the sensitivity bounds which can be
achieved. Nevertheless, at the SLHC, it should be possible to determine
the couplings of a KK gluon resonance with a mass of up to 3~TeV with a
precision of $5-50\%$ in most models.

Our results are subject to a number of uncertainties and thus should be
interpreted with care. Foremost, since most background processes are not
known at NLO, all our signal and background
calculations have been carried out at LO, and thus are subject to 
substantial renormalization and factorization uncertainties. A perhaps
even larger uncertainty originates from the $b$-tagging efficiency and
the light quark and gluon jet misidentification probability at large
$t\bar t$ invariant masses, which is only poorly known at present. PDF
uncertainties, on the other hand, appear to be relatively
small~\cite{Nadolsky:2008zw}. 

The numerical results presented here were obtained by combining
information from the $m(t\bar t)$ and the $p_T(t\to b\ell\nu)$
distributions. Since we ignore correlated systematic uncertainties, our
results are somewhat optimistic. On the other hand, our background
estimate has been deliberately conservative. Furthermore, in future
studies one may include additional distributions in the analysis such as
the transverse momentum distribution of the charged lepton which is
sensitive to the chirality of the coupling of the KK gluon to the top
quark. This could potentially improve the accuracy on the KK gluon
couplings which may be obtained at the LHC and SLHC.

\acknowledgements
We would like to thank J.~Boersma, 
T.~LeCompte, B.~Lillie, and T.~Tait
for useful discussions. One of us would like to thank the
Fermilab Theory Group, where part of this work was done, for
their generous hospitality. This research was supported in part by the
National Science Foundation under grant No.~PHY-0456681 and the
Department of Energy under grant DE-FG02-91ER40685.


\bibliographystyle{plain}

\end{document}